\documentclass[prl,reprint,superscriptaddress,floatfix,longbibliography]{revtex4-2}
\usepackage{graphicx}
\usepackage[colorlinks=true,citecolor=blue]{hyperref}
\usepackage{amsmath}
\usepackage{xcolor}
\usepackage{amssymb}
\usepackage{soul}

\begin{document}

\title{Emergence of Exotic Spin Texture in Supramolecular Metal Complexes on a 2D Superconductor}

\author{Viliam Va\v{n}o}
\affiliation{Department of Applied Physics, Aalto University, FI-00076 Aalto, Finland}

\author{Stefano Reale}
\affiliation{Center for Quantum Nanoscience (QNS), Institute for Basic Science (IBS), Seoul 03760, Republic of Korea}
\affiliation{Ewha Womans University, Seoul 03760, Republic of Korea}
\affiliation{Department of Energy, Politecnico di Milano, Milano 20133, Italy}

\author{Orlando J. Silveira}
\affiliation{Department of Applied Physics, Aalto University, FI-00076 Aalto, Finland}

\author{Danilo Longo}
\affiliation{CIC nanoGUNE-BRTA, 20018 Donostia-San Sebastián, Spain}

\author{Mohammad Amini}
\affiliation{Department of Applied Physics, Aalto University, FI-00076 Aalto, Finland}

\author{Massine Kelai}
\affiliation{Center for Quantum Nanoscience (QNS), Institute for Basic Science (IBS), Seoul 03760, Republic of Korea}
\affiliation{Ewha Womans University, Seoul 03760, Republic of Korea}

\author{Jaehyun Lee}
\affiliation{Center for Quantum Nanoscience (QNS), Institute for Basic Science (IBS), Seoul 03760, Republic of Korea}
\affiliation{Department of Physics, Ewha Womans University, Seoul 03760, Republic of Korea}

\author{Atte Martikainen}
\affiliation{Department of Physics, P.O. Box 4, FI-00014 University of Helsinki, Finland}

\author{Shawulienu Kezilebieke}
\affiliation{Department of Physics, Department of Chemistry and Nanoscience Center, 
University of Jyväskyl\"a, FI-40014 University of Jyväskyl\"a, Finland}

\author{Adam S. Foster}
\affiliation{Department of Applied Physics, Aalto University, FI-00076 Aalto, Finland}
\affiliation{WPI Nano Life Science Institute (WPI-NanoLSI), Kanazawa University, Kakuma-machi, Kanazawa 920-1192, Japan}

\author{Jose L. Lado}
\affiliation{Department of Applied Physics, Aalto University, FI-00076 Aalto, Finland}

\author{Fabio Donati}
\email{Corresponding authors. Email: donati.fabio@qns.science, peter.liljeroth@aalto.fi, lhyan@suda.edu.cn}
\affiliation{Center for Quantum Nanoscience (QNS), Institute for Basic Science (IBS), Seoul 03760, Republic of Korea}
\affiliation{Department of Physics, Ewha Womans University, Seoul 03760, Republic of Korea}

\author{Peter Liljeroth}
\email{Corresponding authors. Email: donati.fabio@qns.science, peter.liljeroth@aalto.fi, lhyan@suda.edu.cn}
\affiliation{Department of Applied Physics, Aalto University, FI-00076 Aalto, Finland}

\author{Linghao Yan}
\email{Corresponding authors. Email: donati.fabio@qns.science, peter.liljeroth@aalto.fi, lhyan@suda.edu.cn}
\affiliation{Institute of Functional Nano \& Soft Materials (FUNSOM), Jiangsu Key Laboratory for Carbon-Based Functional Materials and Devices, Joint International Research Laboratory of Carbon-Based Functional Materials and Devices, Soochow University, Suzhou 215123, China}
\affiliation{Department of Applied Physics, Aalto University, FI-00076 Aalto, Finland}

\begin{abstract}

Designer heterostructures have offered a very powerful strategy to create exotic superconducting states by combining magnetism and superconductivity. In this work, we use a heterostructure platform combining supramolecular metal complexes (SMCs) with a quasi-2D van der Waals (vdW) superconductor NbSe$_2$. Our scanning tunneling microscopy (STM) measurements demonstrate the emergence of Yu-Shiba-Rusinov (YSR) bands arising from the interaction between the SMC magnetism and the NbSe$_2$ superconductivity. Using X-ray absorption spectroscopy (XAS) and X-ray magnetic circular dichroism (XMCD) measurements, we show the presence of antiferromagnetic coupling between the SMC units. These result in the emergence of an unconventional $3\times3$ reconstruction in the magnetic ground state that is directly reflected in real space modulation of the YSR bands. The combination of flexible molecular building blocks, frustrated magnetic textures, and superconductivity in heterostructures establishes a fertile starting point to fabricating tunable quantum materials, including unconventional superconductors and quantum spin liquids.

\end{abstract}

\date{\today}

\maketitle
Supramolecular metal complexes (SMCs) are a highly versatile class of metal-organic materials that combine flexible design, widely tuneable properties and facile access to large scale structures via molecular self-assembly \cite{Dong2016Self-assemblySurfaces}. The ability to tune the lattice geometry, the  magnetism and spin-orbit coupling arising from the metal atoms within these complexes, along with the diverse interactions with the substrate, has led to the prediction of a multitude of extraordinary electronic properties exhibited by SMCs \cite{Springer2020,Jiang2021}. Only a handful of these effects has been realized in practice, including the manifestation of magnetic properties \cite{Gambardella2009,Girovsky2017}, surface-state engineering \cite{RevModPhys.94.045008}, and realizing structures potentially hosting topologically non-trivial band structures \cite{Gao2019}. 

While SMCs can host intrinsic exotic states, creating heterostructures with proximity effects offers even more possibilities. For example, combining magnetic impurities with superconductivity can give rise to Yu-Shiba-Rusinov (YSR) states \cite{Yu1965,Shiba1968,Rusinov1969}. Arranging such impurities in 1D and 2D lattices together with a suitable magnetic texture and spin-orbit interactions provides a route to various artificial topological superconducting states \cite{NadjPerge2014,Rontynen2015,Li2016,Lu2016,Rachel2017,Poyhonen2018,Schneider2022,Flensberg2021,yazdani2023hunting}. Even though YSR bands have been successfully achieved in assemblies of individual magnetic atoms, and with inorganic chains and layers \cite{Ji2010,Heinrich2018,yazdani2023hunting,Schneider2023,bazarnik2023antiferromagnetism,Soldini2023},
the emergence of YSR bands in molecular systems still remains elusive 
\cite{Franke2011,Hatter2015,Mier2021,farinacci2024}. The downside of this approach is that molecular systems on typical bulk superconducting substrates combine the large lattice constants of SMCs with small YSR orbitals. This leads to very limited overlap between the YSR states, hindering the potential creation of YSR bands. In addition, magnetic coupling between the SMC building blocks is typically small.

\begin{figure*}[t!]
    \centering
    \includegraphics[width=.95\textwidth]{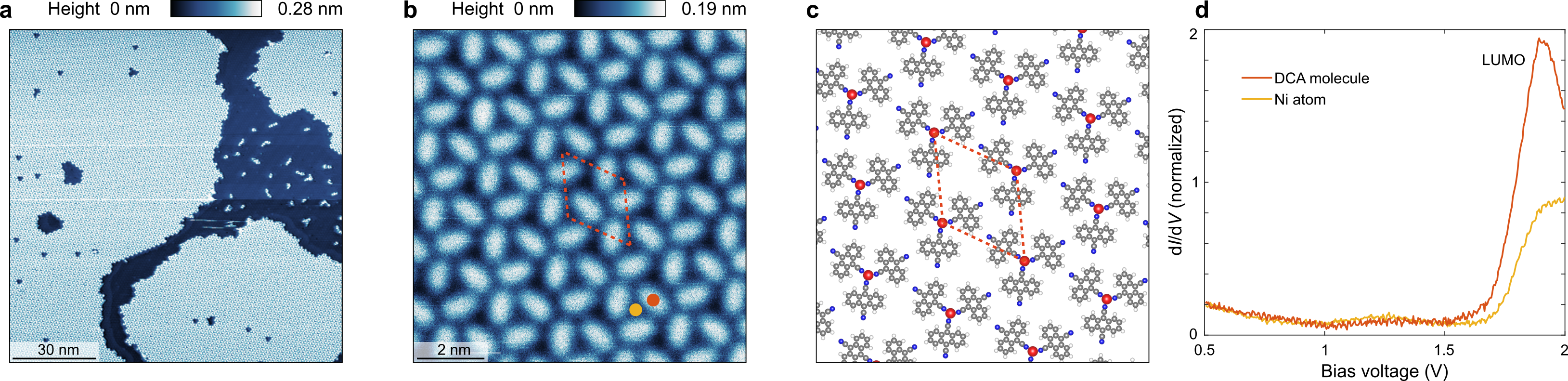}
	\caption{Magnetic NiDCA$_3$ SMC on superconducting NbSe$_2$. (a,b) Large area STM image (a, $V = 1.73$~V and $I = 10$~pA) and small area STM image (b, $V = 1$~V and $I = 5$~pA) of NiDCA$_3$ SMC on NbSe$_2$. (c) DFT calculated structure of the NiDCA$_3$ SMC on the NbSe$_2$ substrate (substrate atoms not shown). (d) d$I$/d$V$ spectra measured on top of the DCA molecule (orange) and Ni atom (yellow) within the NiDCA$_3$ SMC.}
    \label{fig:structure}
\end{figure*}

NbSe$_2$ has been shown to host YSR states with spatially extended orbitals due to its quasi-2D character \cite{Menard2015,Kezilebieke2018,Liebhaber2022}. We demonstrate that this allows for a significant overlap of the YSR states in SMC/NbSe$_2$ heterostructures resulting in the formation of robust YSR bands. Furthermore, we show the presence of antiferromagnetic coupling within the SMC layer by synchrotron-based X-ray absorption spectroscopy (XAS) and magnetic circular dichroism (XMCD). Together with the triangular SMC lattice, this results in magnetic frustration giving rise to an unconventional $3\times3$ reconstruction in the magnetic ground state. This modulation is directly reflected in the local-density of states (LDOS) of the YSR bands that we probe via scanning tunneling microscopy (STM) and spectroscopy (STS). The successful synthesis of magnetically coupled SMC on NbSe$_2$ and the formation of YSR bands thus provides an archetypal platform for engineering unconventional quantum phases in SMC/superconductor hybrid systems. 

\textbf{Heterostructure of NiDCA$_3$ on NbSe$_2$.} The STM image in Fig.~\ref{fig:structure}a shows the results of the growth of a single-layer NiDCA$_3$ (DCA = dicyanoanthracene) SMC on NbSe$_2$ (details of the sample preparation are given in the Supplemental Materials (SM) \cite{SM}). The bright areas are self-assembled SMC islands extending over tens of nanometers, with smaller size SMC clusters randomly distributed between them. A smaller area STM image (Fig.~\ref{fig:structure}b) reveals the triangular lattice of the SMC, which has the structure matching the density functional theory (DFT) calculations shown in Fig.~\ref{fig:structure}c. The building unit of the SMC consists of three DCA molecules surrounding a Ni atom, forming a NiDCA$_3$ single complex arranged in a triangular lattice. This metal coordination is similar to the earlier results on Co-DCA and Cu-DCA networks on graphene and NbSe$_2$ and in a stark contrast with the close-packed assembly of pure DCA molecules on NbSe$_2$ \cite{Kumar2018,Yan2021,Yan2021B}. The magnetic Ni atoms create a triangular lattice of magnetic impurities on the superconducting NbSe$_2$ substrate (Fig.~\ref{fig:structure}c), with an experimentally determined lattice constant of 2.03~nm that matches the value of 2.07~nm calculated by DFT \cite{SM}. The d$I$/d$V$ spectra taken on top of Ni atom and DCA molecules show a broad peak between 1.5~V and 2~V (Fig.~\ref{fig:structure}d), representing the typical metal-ligand bonding orbital features formed in the single complex (see Fig.~S1 in the SM) \cite{Liljeroth2010,Kumar2018}. 

\begin{figure}[b!]
    \centering
    \includegraphics[width=.99\columnwidth]{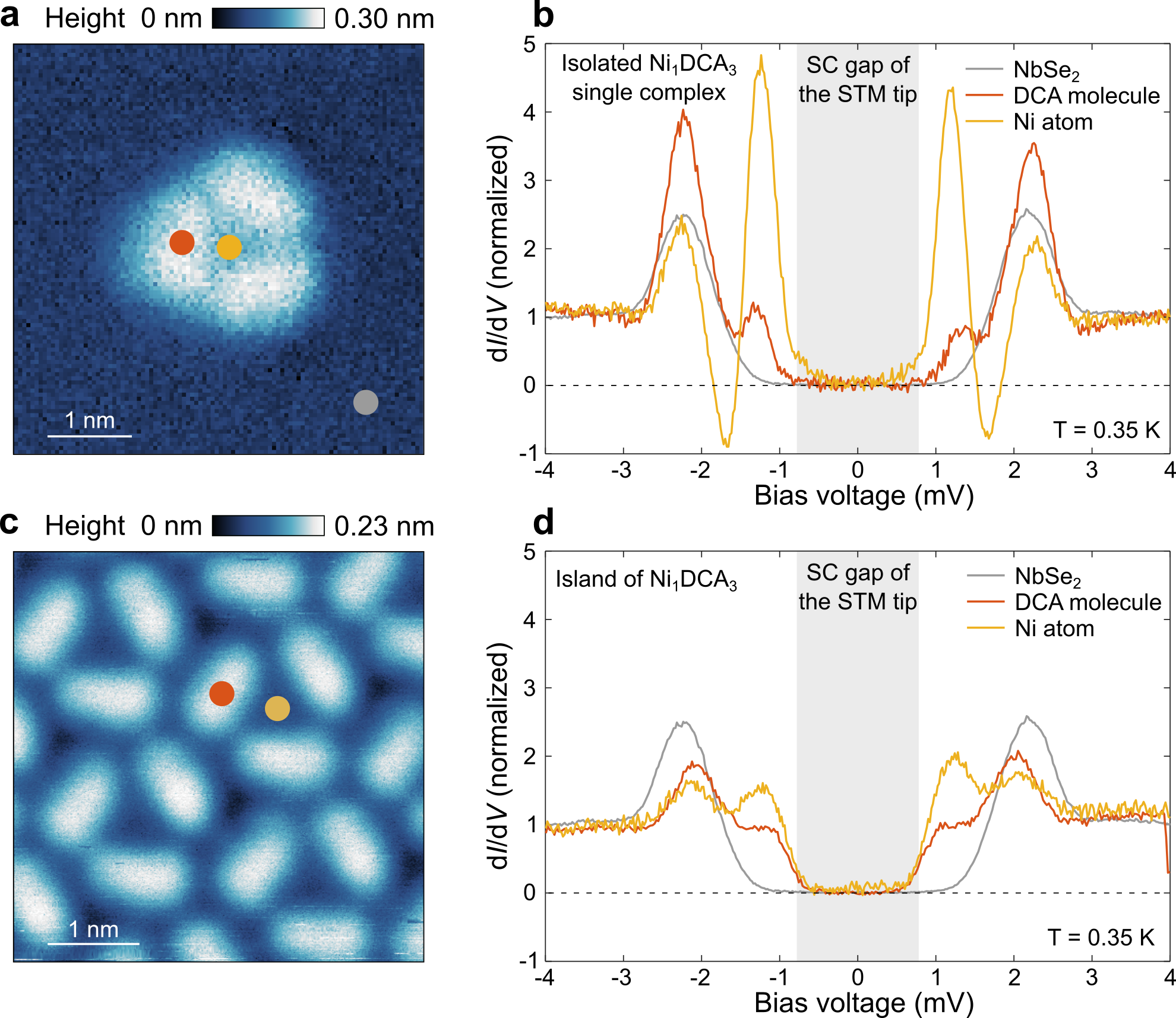}
	\caption{Evolution from a single magnetic impurity to a lattice of magnetic impurities. (a) STM image of isolated NiDCA$_3$ single complex ($V = 1$~V and $I$ = $5$~pA). (b) d$I$/d$V$ spectra measured with a superconducting tip on NbSe$_2$ (gray) and on Ni atom (yellow) and DCA molecule (orange) of the isolated NiDCA$_3$ single complex. (c) STM image of NiDCA$_3$ SMC on NbSe$_2$ ($V = 1$~V and $I$ = $10$~pA). (d) d$I$/d$V$ spectra measured with superconducting NbSe$_2$ tip on NbSe$_2$ (gray) and on the Ni (yellow) and DCA (orange) within the NiDCA$_3$ SMC.}
    \label{fig:spectra}
\end{figure}

\begin{figure*}[t]
    \centering
    \includegraphics{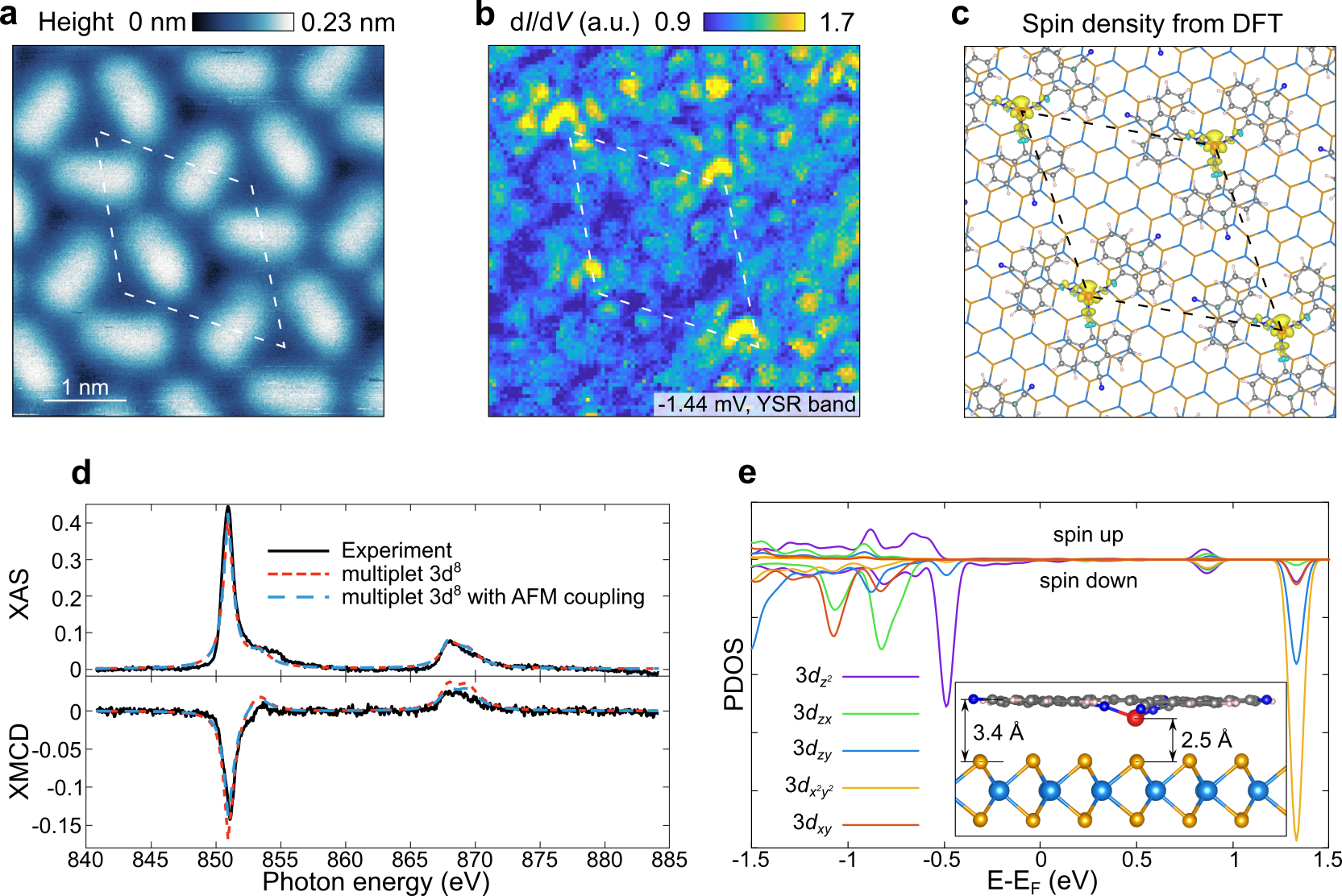}
	\caption{Spatial distribution of the YSR bands and electronic configuration of the Ni. (a) STM image of NiDCA$_3$ SMC on NbSe$_2$ ($V = 1$~V and $I$ = $10$~pA). (b) d$I$/d$V$ map of the area shown in (a) at $V = -1.44$ mV, corresponding to the energy of the YSR band (feedback parameters $V = 4$~mV, $I$ = $80$~pA and $V_\mathrm{mod} = 100$ $\mu$V). (c) Spatial distribution of the spin density of NiDCA$_3$/NbSe$_2$, calculated by DFT. (d) XAS and corresponding XMCD of the NiDCA$_3$/NbSe$_2$ at the Ni $L_{2,3}$ edges, acquired at normal incidence. Experiments (blue lines) are compared with multiplet calculations performed for a 3d$^8$ configuration neglecting (green dashed lines) or including (red dashed lines) a magnetic coupling among the Ni centers, see text for more details ($T_{\rm{sample}} = 1.9$~K, magnetic field $B = 6.0$~T). (e) PDOS of Ni 3d orbitals within the NiDCA$_3$ SMC, calculated by DFT. The structure is shown in the inset, where the light blue spheres correspond to Nb atoms, yellow to Se atoms, red to Ni atom, grey to C atoms, blue to N atoms and white to H atoms.}

    \label{fig:orbital}
\end{figure*}
\textbf{Emergence of Yu-Shiba-Rusinov bands.} A single magnetic impurity on top of a superconductor induces a pair of YSR states inside the superconducting gap \cite{Yu1965,Shiba1968,Rusinov1969,Heinrich2018}. Analogous to the electronic lattice, the overlap of YSR states leads to their hybridization and thus to the formation of YSR bands. These appear as broad features in the spectral function rather than the sharp peaks stemming from single states. We investigated this using STM with a superconducting NbSe$_2$ tip that improves the energy resolution of the experiment and more importantly, the stability of the measurements on the molecular assemblies (see SM for details). Our experimental observations on SMCs on NbSe$_2$ confirm the expected transition from single YSR states to YSR bands: when an isolated single complex is positioned on NbSe$_2$ (Fig.~\ref{fig:spectra}a), two sharp peaks appear in the tunneling spectroscopy symmetrically around the Fermi level (Fig.~\ref{fig:spectra}b). As these single complexes form a SMC lattice (Fig.~\ref{fig:spectra}c), the sharp peaks in the tunneling spectrum become broader and weaker in intensity (Fig.~\ref{fig:spectra}d). This is a sign of strong hybridization and a band formation, which is enabled by the large spatial extent of YSR states of NbSe$_2$ \cite{Menard2015,Kezilebieke2018,Liebhaber2022} and leads to a significant overlap of these states between neighbouring single complexes. Another sign of YSR band formation, as we will show in the following, is that there is a strong YSR LDOS across the whole unit cell of the SMC. Alternatively, the observed change in the YSR states could be due to the change in the effective temperature of the instrument or due to Pauli depairing. Both of these scenarios lead to changes in the spectroscopy of the superconducting gap, which we do not observe.

One of the advantages of our work, stemming from the large unit cell and extended YSR orbitals, is the further insight into spatial behavior of the YSR bands when probed by STM. This can reveal information about the orbital nature of the electronic state that induces the YSR state. While numerous studies have reported this for coupled YSR dimers and chains, research efforts focused on 2D YSR lattices remain notably limited \cite{bazarnik2023antiferromagnetism,Soldini2023}. The LDOS map in Fig.~\ref{fig:orbital}b visualizes the YSR band, which was measured on the area shown in Fig.~\ref{fig:orbital}a.  The intensity of the YSR bands is particularly concentrated around the central Ni atom of each single complex, where DFT finds maximal spin density. However, due to the remarkable spatial extent of the YSR states, a pronounced intensity pattern is visible across the whole unit cell even where the spin density vanishes (Figs.~\ref{fig:orbital}c). The additional long-range modulation of the YSR band occurring over several unit cells shown in Fig.~\ref{fig:orbital}a is due to a magnetic ground state (see below).

\textbf{Magnetism of Ni atoms from XAS and XMCD.} To gain insight into the electronic configuration of the Ni atoms in the SMC, XAS and XMCD measurements were conducted at the DEIMOS beamline of the SOLEIL synchrotron (details in the SM). Figure~\ref{fig:orbital}d shows the spectra collected at the $L_{2,3}$ edges of Ni exhibiting a characteristic $3d^8$ lineshape \cite{Klewe2014}. The large XMCD signal indicates a high spin configuration with a triplet $S = 1$ ground state. Angular-dependent XAS and XMCD measurements reveal a weak magnetic anisotropy, with a greater magnetization along the out-of-plane direction (see SM). To quantitatively analyze the magnetic state of the Ni atoms, we performed a fitting procedure using simulated spectra based on atomic multiplet calculations. The calculations considered a Ni $3d^8$ configuration, accounting for intra-atomic electron-electron interactions, spin-orbit coupling, crystal field effects within the C$_{3v}$ symmetry, and an external magnetic field (see SM). The result of the fit, overlaid to the experiment in Fig.~\ref{fig:orbital}d, supports the inferred $3d^8$ configuration with a $S = 1$ triplet ground state and allows quantifying a value for the magnetic anisotropy of about 0.1~meV. While the multiplet model nicely reproduces the shape of both XAS and XMCD spectra, the amplitude of the XMCD signal appears to be slightly overestimated. As discussed in the following section, the inclusion of antiferromagnetic interactions between the Ni atoms enables a comprehensive reproduction of the XMCD amplitude at all magnetic fields and temperatures. Additionally, we computed the projected density of states (PDOS) for the Ni d-orbitals (Fig.~\ref{fig:orbital}e) by DFT and determined the orbital occupation of $3d^8$ magnetic state for the Ni atoms based on inputs from X-ray absorption spectroscopy (Fig.~\ref{fig:orbital}d). These results demonstrate that the studied SMC realizes a $S = 1$ lattice.

\begin{figure}[t!]
    \centering
    \includegraphics[width=.95\columnwidth]{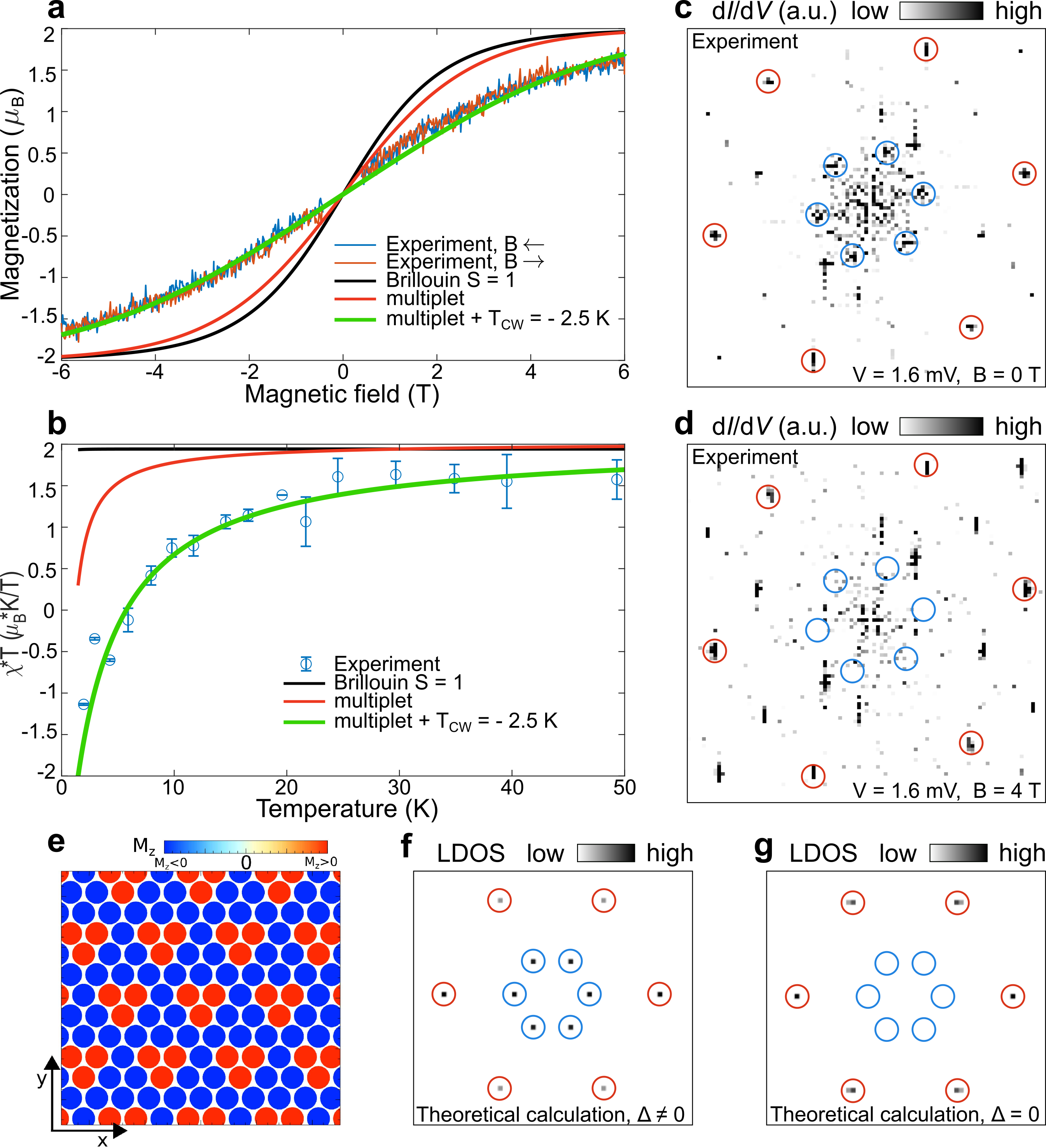}
	\caption{Magnetic order of the NiDCA$_3$ SMC. (a) Hysteresis curve at $T = 2.0$~K obtained by measuring the XMCD signal while sweeping the magnetic field. Brillouin curve for $S=1$ (solid black line), magnetization from the multiplet model (red solid line) and including antiferromagnetic interaction with a Curie-Weiss temperature $T_{CW} = -2.5$~ K (green solid line) are shown for comparison. (b) Temperature dependence of the susceptibility obtained from hysteresis curves at various temperatures and comparison with the related models. Data shown in (a) and (b) were acquired at normal incidence. (c,d) d$I$/d$V$($\vec{q}$,V) measured at the energy of the YSR bands ($V = 1.6$ mV) at 0 Tesla (c) and 4 Tesla (d), acquired at $T = 350$ mK. Red circles correspond to the Bragg peaks and blue circles to 3 $\times$ 3 peaks related to the anti-ferromagnetic order. (e) Illustration of the $3\times 3$ Ising-like magnetic ground state consistent with our observations. (f) Simulated LDOS($\vec{q}$) of the in-gap Yu-Shiba-Rusinov states, allowing the imaging of the magnetic reconstruction in the YSR bandstructure. (g) In the absence of superconductivity, the additional peaks are not observed in the LDOS($\vec{q}$).
 }
 
    \label{fig:magnetism}
\end{figure}

To unravel the nature and strength of the magnetic interactions in the SMC, we acquired magnetization loops by measuring the XMCD signal (Fig.~\ref{fig:orbital}d, down) while sweeping the applied magnetic field over a range between -6 and +6 T, and at temperatures from 2 to 50 K. The curve acquired at 2 K shown in Fig.~\ref{fig:magnetism}a shows a smooth reversible loop, with no hysteresis or magnetization jumps between the backward and forward branch indicating the absence of stable magnetic structures at that temperature. However, the amplitude and slope of the curve are lower than the expected values of a $S = 1$ Brillouin function at $T = 2.0$~K (black solid line). Including the full multiplet structure (red solid line) only marginally improves the discrepancy between experiment and theory. The same type of discrepancy appears in the magnetic susceptibility extracted from the magnetization loops as a function of temperature (see Fig.~\ref{fig:magnetism}b), with both $S = 1$ Brillouin and full multiplet model not capturing the experimental trend. The model improves remarkably by assuming the presence of antiferromagnetic interactions through $n$ nearest-neighbor pairwise exchange $J^\mathrm{ex}$ resulting in a Curie-Weiss temperature $T_\mathrm{CW} = -\frac{S(S+1)}{3k_\mathrm{B}}nJ^\mathrm{ex}$ \cite{Johnston2012}, which we take as a free parameter of the fit (green solid line). The combination of multiplet calculations and the just mentioned antiferromagnetic interaction provides an excellent agreement with the experiment for a $T_{CW} = -2.5 \pm 0.4$~K, which corresponds to an exchange energy of about $0.2$ meV. This indicates that the exchange interaction between the Ni atoms dominates over the magnetic anisotropy. The relation between $T_{CW}$ and the 
ordering N\'eel temperature $T_N$ depends on the geometry of the spin lattice and on the neighbor order contributing to the magnetic coupling.  For the case of a frustrated triangular-type lattice where only the first nearest neighbors are considered ~\cite{Johnston2012}, $T_\mathrm{N} = -T_\mathrm{CW}/2 = 1.2 \pm 0.2$~K, i.e., lower than the base temperature for XMCD measurements, in line with what was observed above.  The low Curie-Weiss and N\'eel temperatures are consistent with the relatively large distances and absence of chemical bonds between the single complexes.

\textbf{$3\times3$ magnetic order from STM measurements.} The combination of antiferromagnetic coupling and a frustrated triangular lattice can give rise to exotic magnetic ground states, such as a quantum spin liquid or 120$^{\circ}$ N\'eel order. To determine the ground state of the SMC, we map the periodicity of the YSR signal with STM at a temperature of 350 mK, i.e. below the expected ordering temperature of 1.25K. The magnetic order can be imprinted in the YSR states as the energy and density distribution of YSR states depend on the relative angle between magnetic moments. The STM results in Figs.~\ref{fig:magnetism}c,d show d$I$/d$V$($\vec{q}$,V), an FFT of a LDOS map at the energy of the YSR bands. The main features here are the Bragg peaks (red circles) and $3\times3$ peaks (blue circles). Within the $\vec{q}$-resolution of the d$I$/d$V$($\vec{q}$,V), our data suggests that the $3\times3$ peaks are aligned with the Bragg peaks. Upon applying an external magnetic field of $B=4$~T, at which there is no observable superconducting gap in the tunneling spectroscopy \cite{Kezilebieke2020}, the Bragg peaks remain while the $3\times3$ peaks disappear. This suggests that the $3\times3$ peaks are connected to the YSR bands that rely on superconductivity. Alternatively, a magnetic field of 4T could align the spins into a ferromagnetic structure, destroying the $3\times3$ supercell. Both of these explanations support the magnetic ground state reconstruction that is imprinted in the YSR bands. The $3\times3$ supercell of YSR bands could also arise due to a moir\'e pattern, where the registry of Ni atoms with respect to Se atoms of NbSe$_2$ repeat on a $3\times3$ supercell. Similarly, the pattern could appear due to a repeating registry between the Ni atoms and $3\times3$ CDW of NbSe$_2$. However, our analysis shows that these cases can be ruled out (see SM for details). Therefore, the supercell is likely to arise from a $3\times3$ magnetic order of the Ni atoms within the SMC. 

Besides, there are additional peaks in the d$I$/d$V$($\vec{q}$,V), and even though we cannot determine their exact origin at this point, they are not related to the superconducting state as they appear the same at 0~T and 4~T.

\textbf{Theoretical modeling.} To explain how the magnetic reconstruction can be observed in the STM signal, we present in Figs.~\ref{fig:magnetism}e-g model calculations for a system with magnetic reconstruction shown in Fig.~\ref{fig:magnetism}e. The magnetic ordering induces an exchange coupling in the underlying substrate, leading to a modulated exchange profile (details shown in the SM). This phenomenology is illustrated here with an Ising-like $3\times 3$ magnetic configuration, realizing a minimal ground state with strong out-of-plane anisotropy. In the case of weaker out-of-plane anisotropy, non-collinear configurations featuring $3\times 3$ periodicity would be consistent with our observations as analyzed in the SM. The magnetic modulation translates into the modulation of the energy of YSR states, reflecting the broken translational symmetry of the magnetic structure. The energy modulation of the YSR states can be visualized as a LDOS modulation at an in-gap energy featuring YSR modes (Fig.~\ref{fig:magnetism}f). In stark contrast, in the absence of superconductivity or a magnetic superstructure, no additional peaks are observed in the FFT of the d$I$/d$V$($\vec{q}$,V) signal (Fig.~\ref{fig:magnetism}g, see SM for details). 
While the STM experiments do not directly probe the magnetism of the SMC, the interplay between magnetism and superconductivity translates the magnetic texture into a spatial modulation of the YSR bands that can be seen in the STM experiments. This allows visualizing frustrated magnetic orders directly in STM and STS experiments.

\textbf{Conclusions.} We have successfully synthesized a magnetic SMC on a quasi-2D superconductor NbSe$_2$, and investigated its electronic and magnetic properties using STM and XMCD. The STM experiments demonstrate that this heterostructure of a magnetic lattice and a superconductor leads to the emergence of YSR bands spanning the entire unit cell of the SMC. Furthermore, combining XMCD and STM results, we discovered an exotic $3\times3$ antiferromagnetic order of the SMC. We use STM in a novel and unique way to extract information about the spin texture of the system, which is imprinted in the YSR bands. These results open up an extremely tunable path towards realizing novel quantum phases, such as topological superconductivity or quantum spin liquids using SMC/superconductor heterostructures. Here, the tunability stems from the choice of metal atoms, molecules and the substrate, which allows for a combination of different ingredients for designing  a wide range of physical systems. 

\textbf{Data availability. }The DFT and multiplet computational data and metadata are freely available under a CC BY 4.0 license on the following link: doi.org/10.5281/zenodo.13969023. The low energy model code is available on https://github.com/joselado/pyqula. Experimental data is available upon reasonable request.

\begin{acknowledgements}
This research made use of the Aalto Nanomicroscopy Center (Aalto NMC) facilities and was supported by the European Research Council (ERC-2017-AdG no.~788185 ``Artificial Designer Materials'') and Academy of Finland (Academy professor funding nos.~318995 and 320555, Academy research fellow nos.~331342, 336243 and nos.~338478 and 346654). L.Y. acknowledges support from the Jiangsu Specially-Appointed Professors Program, Natural Science Foundation of Jiangsu Province (Grants No BK20230474), Suzhou Key Laboratory of Surface and Interface Intelligent Matter (Grant SZS2022011), Gusu Innovation and Entrepreneurship Talent Program - Major Innovation Team (ZXD2023002), Suzhou Key Laboratory of Functional Nano \& Soft Materials, Collaborative Innovation Center of Suzhou Nano Science \& Technology, and the 111 Project. Computing resources from the Aalto Science-IT project and CSC, Helsinki are gratefully acknowledged. A.S.F has been supported by the World Premier International Research Center Initiative (WPI), MEXT, Japan. S.R, M.K, J.L, and F.D. acknowledge support from the Institute for Basic Science under Grant IBS-R027-D1. D.L. acknowledges support from the MSCA project (project number: 101064332). We are grateful to Fadi Choueikani for assistance with XAS and XMCD experiments and to the SOLEIL staff for smoothly running the facility. The authors would like to thank Bernard Muller for his significant contribution to the construction of DEIMOS beamline (design and engineering) and Florian Leduc for his support. The MBE chamber used during the XAS/XCMD experiment on DEIMOS has been funded by the Agence National de la Recherche; grant ANR-05-NANO-073.
\end{acknowledgements}

\bibliography{Refs}

\end{document}


\title{Supplemental Material: Emergence of Exotic Spin Texture in Supramolecular Metal Complexes on a 2D Superconductor}

\author{Viliam Va\v{n}o}
\affiliation{Department of Applied Physics, Aalto University, FI-00076 Aalto, Finland}

\author{Stefano Reale}
\affiliation{Center for Quantum Nanoscience (QNS), Institute for Basic Science (IBS), Seoul 03760, Republic of Korea}
\affiliation{Ewha Womans University, Seoul 03760, Republic of Korea}
\affiliation{Department of Energy, Politecnico di Milano, Milano 20133, Italy}

\author{Orlando J. Silveira}
\affiliation{Department of Applied Physics, Aalto University, FI-00076 Aalto, Finland}

\author{Danilo Longo}
\affiliation{CIC nanoGUNE-BRTA, 20018 Donostia-San Sebastián, Spain}

\author{Mohammad Amini}
\affiliation{Department of Applied Physics, Aalto University, FI-00076 Aalto, Finland}

\author{Massine Kelai}
\affiliation{Center for Quantum Nanoscience (QNS), Institute for Basic Science (IBS), Seoul 03760, Republic of Korea}
\affiliation{Ewha Womans University, Seoul 03760, Republic of Korea}

\author{Jaehyun Lee}
\affiliation{Center for Quantum Nanoscience (QNS), Institute for Basic Science (IBS), Seoul 03760, Republic of Korea}
\affiliation{Department of Physics, Ewha Womans University, Seoul 03760, Republic of Korea}

\author{Atte Martikainen}
\affiliation{Department of Physics, P.O. Box 4, FI-00014 University of Helsinki, Finland}

\author{Shawulienu Kezilebieke}
\affiliation{Department of Physics, Department of Chemistry and Nanoscience Center, 
University of Jyväskyl\"a, FI-40014 University of Jyväskyl\"a, Finland}

\author{Adam S. Foster}
\affiliation{Department of Applied Physics, Aalto University, FI-00076 Aalto, Finland}
\affiliation{WPI Nano Life Science Institute (WPI-NanoLSI), Kanazawa University, Kakuma-machi, Kanazawa 920-1192, Japan}

\author{Jose L. Lado}
\affiliation{Department of Applied Physics, Aalto University, FI-00076 Aalto, Finland}

\author{Fabio Donati}
\email{Corresponding authors. Email: donati.fabio@qns.science, peter.liljeroth@aalto.fi, lhyan@suda.edu.cn}
\affiliation{Center for Quantum Nanoscience (QNS), Institute for Basic Science (IBS), Seoul 03760, Republic of Korea}
\affiliation{Department of Physics, Ewha Womans University, Seoul 03760, Republic of Korea}

\author{Peter Liljeroth}
\email{Corresponding authors. Email: donati.fabio@qns.science, peter.liljeroth@aalto.fi, lhyan@suda.edu.cn}
\affiliation{Department of Applied Physics, Aalto University, FI-00076 Aalto, Finland}

\author{Linghao Yan}
\email{Corresponding authors. Email: donati.fabio@qns.science, peter.liljeroth@aalto.fi, lhyan@suda.edu.cn}
\affiliation{Institute of Functional Nano \& Soft Materials (FUNSOM), Jiangsu Key Laboratory for Carbon-Based Functional Materials and Devices, Joint International Research Laboratory of Carbon-Based Functional Materials and Devices, Soochow University, Suzhou 215123, China}
\affiliation{Department of Applied Physics, Aalto University, FI-00076 Aalto, Finland}


\maketitle

\section*{Methods}
\textbf{Sample preparation.} 
Sample preparation and STM experiments were carried out in ultrahigh vacuum (UHV) systems with a base pressure of $\sim 10^{-10}$~mbar. The 2H–NbSe$_2$ single crystal (HQ Graphene, the Netherlands) was cleaved \emph{in-situ} in the vacuum. The supramolecular NiDCA$_3$ complexes were fabricated by the sequential deposition of 9,10-dicyanoanthracene (DCA, Sigma Aldrich) molecules (evaporation temperature 100 $^\circ$C) and Ni atoms onto the NbSe$_2$ substrate held at room temperature. Further annealing the sample at 40$^\circ$C - 70 $^\circ$C results in larger self-assembled SMC islands. We initially explored a number of transition metal (Ni, Fe, Mn) systems. Among these, Ni-DCA samples demonstrated superior growth characteristics, forming large, well-ordered areas.

\textbf{STM measurements.} After preparation the sample was inserted into the low-temperature STM housed in the same UHV system and subsequent experiments were performed at $T = 5$~K (CreaTec LT-STM) and $T = 350$~mK (Unisoku USM-1300). STM images were taken in the constant-current mode. d$I$/d$V$ spectra were recorded by standard lock-in detection while sweeping the sample bias in an open feedback loop configuration, with a peak-to-peak bias modulation specified for each measurement and at a frequency of 911~Hz. Most of the experiments were carried out with NbSe$_2$ coated STM tips prepared as described previously \cite{Kezilebieke2018}. STM measurements of the SMC on NbSe$_2$ are extremely challenging due to weak interactions within the SMC, often causing the STM tip to pick up a single molecule. Moreover, the SMC interacts more strongly with the STM tip than with the NbSe$_2$ substrate with a van der Waals character. Due to this, we had low success rate carrying STM experiments at the energies of the superconducting gap even at low currents.

\textbf{XAS and XMCD measurements.} Circular and linear XAS measurements of the SMC on NbSe$_2$ were performed on the DEIMOS beamline \cite{Ohresser2014,Joly2014}  at the SOLEIL synchrotron, France (proposals number 20210238 and 20220372). The sample was prepared \emph{in} situ  by following the same procedure described above  and characterized prior to the X-ray experiments using an Omicron VT-STM available at the beamline. The sample was transferred in UHV ($< 1 \times 10^{-9}$~mbar) from the preparation chamber to the XAS measurement stage without breaking the vacuum. The measurements were performed using the V$^2$TI variable temperature insert (lowest temperature 1.8 K) in the temperature range between 2~K and 50~K. The accuracy of the temperature reading was verified by acquiring magnetization loops over an ensemble of paramagnetic neodymium atoms on the same NbSe$_2$ crystals used for our experiments. Fitting with a Brillouin function gives a discrepancy of 0.4 K from the reading of the temperature sensor. This discrepancy is in line with previous reports \cite{malavolti2018,serrano2022,poggini2024} and represent the largest uncertainty on the inferred values of Curie-Weiss temperature provided in the text. Magnetic fields of up to 6.0~T were applied parallel to the X-ray beam. 
Circularly and linearly polarized light from the synchrotron source was directed on the sample with the photon beam either at normal or at grazing (60°) incidence. The circular right (CR) and left (CL) photon polarizations are defined with respect to the photon beam direction while the linear horizontal (LH) and vertical (LV) photon polarizations are defined with respect to the sample surface. All signals have been acquired in total electron yield and normalized at the value of a mesh current placed upstream in the beam line to remove time and energy-related variation of the photon intensity. In addition, all spectra have been normalized to the related pre-edge intensity. Background spectra over the energy range of the Ni $L_{2,3}$ edges were acquired on a clean NbSe$_2$, normalized to the absorption at the related pre-edge energy, and subtracted from the spectra of sample covered with the SMC. To compare with theory, we normalize both experiment and simulated spectra from multiplet calculation to the integral of the corresponding total XAS.
Magnetization loops were acquired by measuring the CR and CL signals at the energy corresponding to maximum absolute value of the XMCD as a function of the magnetic field and obtaining the resulting curve by subtracting CR-CL. The amplitude of the curve was normalized such that the value of $M$ at $T = 2$~K and $B=6$~T reflects the total magnetic moment obtained from sum rules in normal incidence, i.e. $m = m_S + m_L$. For the shown sample and assuming a $3d^8$ configuration, we find a spin magnetic moment $m_S = 1.32 \mu_B$, an orbital magnetic moment $m_L = 0.24~\mu_B$, and $m = 1.56~ \mu_B$. Curves measured at higher temperature have been rescaled keeping the ratio with the curve measured at $T = 2$~ K.
The magnetic susceptibility has been estimated by fitting each magnetization loop with a Brillouin function and taking the slope of the fitted curve at $B = 0$~T.

\textbf{Multiplet calculations.} Spectra of Ni $L_{2,3}$ edges were simulated using the Quanty multiplet code \cite{Haverkort2016} including electron-electron interaction, spin-orbit coupling, crystal field splitting of the 3d orbitals, and Zeeman interaction. Values of the Slater integrals and spin-orbit coupling were obtained using atomic calculations with the Cowan code \cite{Cowan1981}. Rescaling factors for the 3d-3d and 2p-3d Slater integrals, as well as the relative on-site energy of the 3d orbitals were used as free parameters to fit the experiments. The fit was performed using a Bayesian optimization algorithm minimizing the sum of the individual mean square error over the set of angular dependent circular and linear absorption spectra. In addition, from the quantum states generated by the multiplet we infer anisotropy parameter $D$ and electron spin $g$-factor that were used as input for a spin Hamiltonian used to fit the magnetization loop at 2~K and magnetic susceptibility data. We used 2 independent parameters for the electron-electron interaction, 3 independent parameters for the onsite orbital energy, plus an additional parameter to account for the magnetic coupling, see SI for more details. The spectra were calculated taking the imaginary part of the Green functions of the electric dipole transition operator applied to the nickel electron wavefunctions with a Lorentzian broadening of 800~meV and a Gaussian broadening of 200~meV. To reproduce the energy dependent life-time of the X-ray absorption excitation \cite{Krause1979} we further applied an additional linear photon energy-dependent Lorentzian broadening ranging from 0.1~eV to 1.2~eV over the Ni $L_3$ and from 0.6~eV to 2.2~eV over the $L_2$ edge. The best values obtained from the fit are shown in Table S1.

\textbf{DFT calculations.} First principles calculations were performed using the spin-polarized DFT as implemented as in the periodic plane-wave basis code QUANTUM-ESPRESSO \cite{Giannozzi_2009}. Van der Waals interactions were included by considering the DFT-D3 functional \cite{10.1063/1.3382344} with a kinetic energy cutoff of 60 Ry, and core electrons were represented by ultra-soft pseudopotentials generated with the Rappe–Rabe–Kaxiras–Joannopoulos (RRKJ) recipe \cite{PhysRevB.41.1227}. A Hubbard $U=3$~eV term was introduced to improve the description of the correlation of the Ni 3d electrons within the simplified version of Cococcioni and de Gironcoli implemented in QUANTUM-ESPRESSO \cite{PhysRevB.71.035105}.

\section{$\mathbf{d}I$/$\mathbf{d}V$ map of the SMC orbital state}
\begin{figure*}[t!]
    \centering
    \includegraphics[width=.7\textwidth]{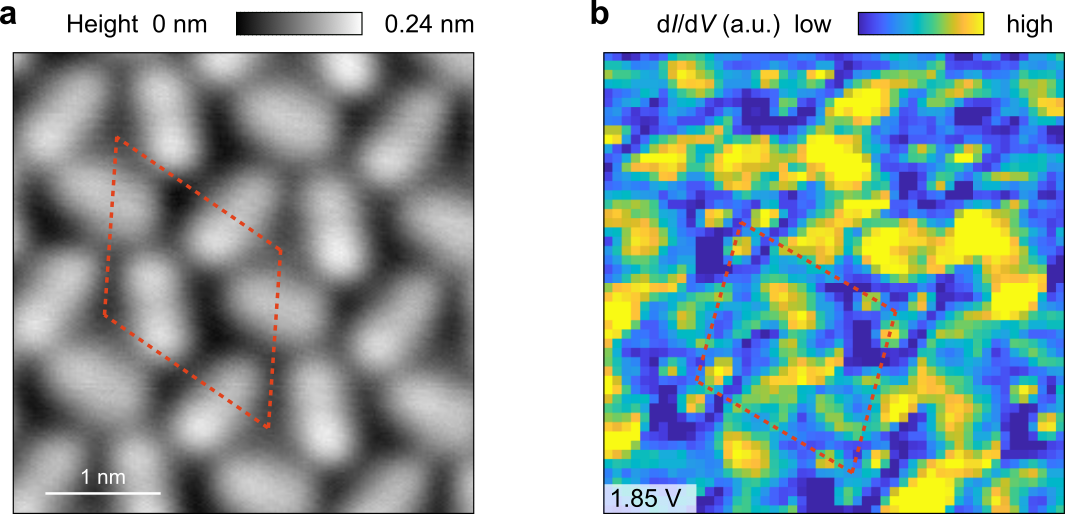}
    \caption{(a) STM image of NiDCA$_3$ SMC on NbSe$_2$ (set point parameters $V = 1$ V and $I$ = $10$ pA). (b) d$I$/d$V$ map of the area shown in (a) at $V = 1.85$ V, corresponding to the energy of the metal-ligand bonding orbital.}
    \label{fig:orbitalmap}
\end{figure*}
The d$I$/d$V$ map at the energy of the orbital level (Fig.~\ref{fig:orbitalmap}b) shows strong contrast at each end of the long axis of the DCA molecule, which is consistent with the appearance of the lowest unoccupied molecular orbital (LUMO) of the DCA molecule. Meanwhile, there is strong contrast near the Ni atom and the nearby cyano (CN) group of DCA molecules, which represents the typical metal-ligand bonding orbital features formed in metal-DCA complexes \cite{Liljeroth2010,Kumar2018}. The d$I$/d$V$ map includes a considerable amount of thermal drift and thus, does not exactly match the STM image of the same area (Fig.~\ref{fig:orbitalmap}a) captured prior to the d$I$/d$V$ map.
  
\section{Real-space $\mathbf{d}I$/$\mathbf{d}V$ maps of the YSR band}
\begin{figure*}[t!]
    \centering
    \includegraphics[width=.9\textwidth]{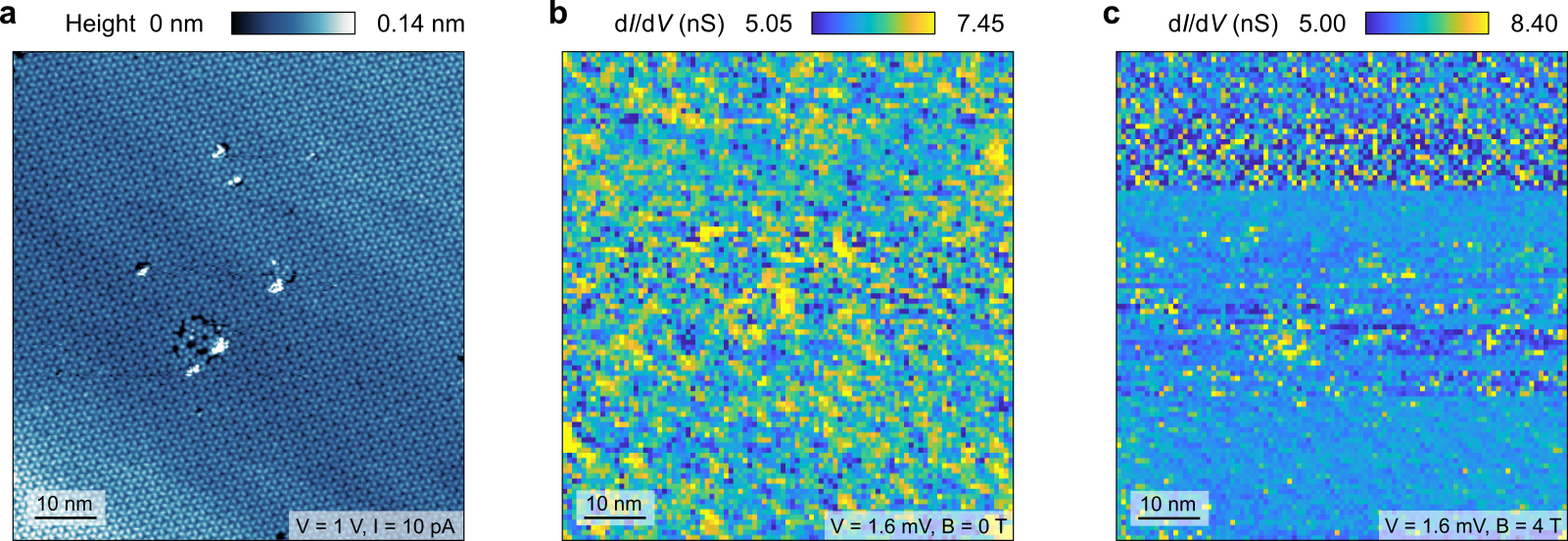}
    \caption{(a) STM image and (b) real-space d$I$/d$V$ map corresponding to the FFT shown in Fig.~4c (set point parameters $V = 1.6$~mV and $I$ = $10$~pA). (c) Real-space d$I$/d$V$ map corresponding to the FFT shown in Fig.~4d (set point parameters $V = 1.6$~mV and $I$ = $10$~pA). }
    \label{fig:real_space}
\end{figure*}
We designed the d$I$/d$V$ maps shown in Fig.~\ref{fig:real_space} for the reciprocal space experiments with minimal spatial resolution, sufficient to capture the first Brillouin zone (at least 2 points per unit cell). Consequently, the reciprocal-space images shown in Fig.~4c,d can be used to identify the Bragg peaks as well as the 3 $\times$ 3 peaks of the YSR bands. While we propose that these peaks arise from magnetic order, it could be possible that they would correspond to a moiré pattern arising between the hexagonal lattice of the NiDCA$_3$ SMC and a rectangular measurement grid. However, such a moiré pattern would result in a stripe pattern, only a severe distortion of the rectangular lattice into a hexagonal lattice could produce a hexagonal moiré pattern and reproduce the 3 $\times$ 3 peaks. Moreover, such pattern would not disappear in magnetic field.

The d$I$/d$V$ map shown in Fig.~\ref{fig:real_space}c includes a tip change. Despite this, the reciprocal space images in Fig.4c,d match perfectly in terms of Bragg peaks and all the other peaks. The only difference is absence or presence of the the 3 $\times$ 3 peaks, which is very unlikely to result from only a tip change.

\begin{figure*}[h!]
    \centering
    \includegraphics[width=.9\textwidth]{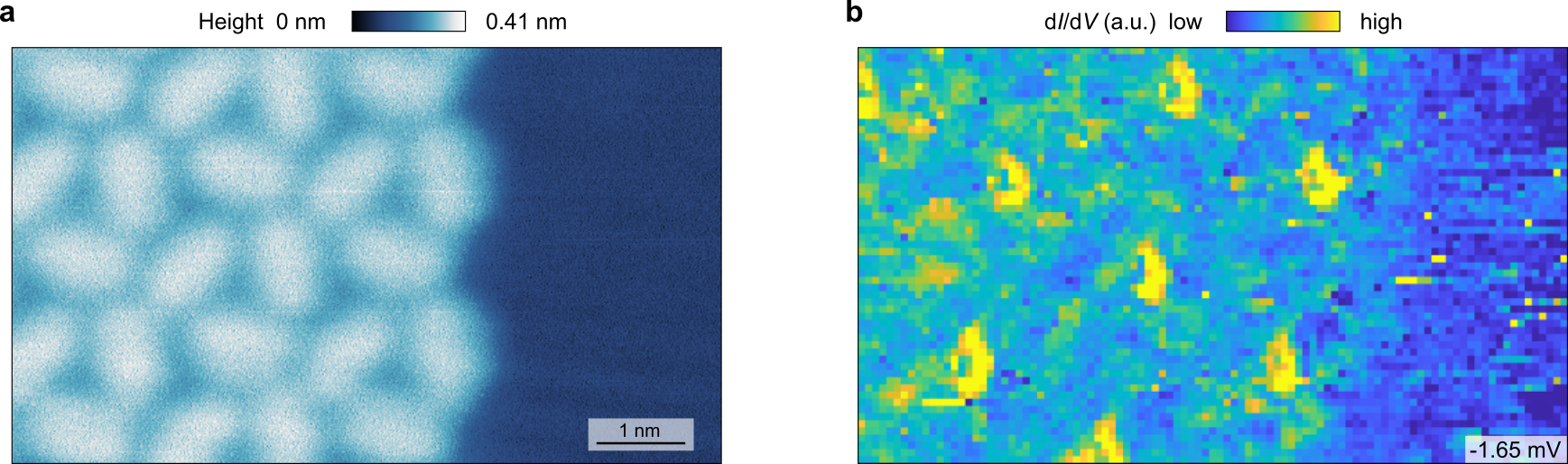}
    \caption{STM image (a) and the corresponding d$I$/d$V$ map (b) measured at the edge of an SMC island. The setpoint parameters for the STM image are $V = 1$~V and $I$ = $10$~pA and for the d$I$/d$V$ map $V = 4$~mV and $I$ = $80$~pA.}
    \label{fig:edge}
\end{figure*}

In addition, we also obtained a high-resolution d$I$/d$V$ map at the edge of an SMC island shown in Fig.~\ref{fig:edge}, where we do not observe any significant changes of the YSR bands at the edge.

\begin{figure*}[h!]
    \centering
    \includegraphics[width=.75\textwidth]{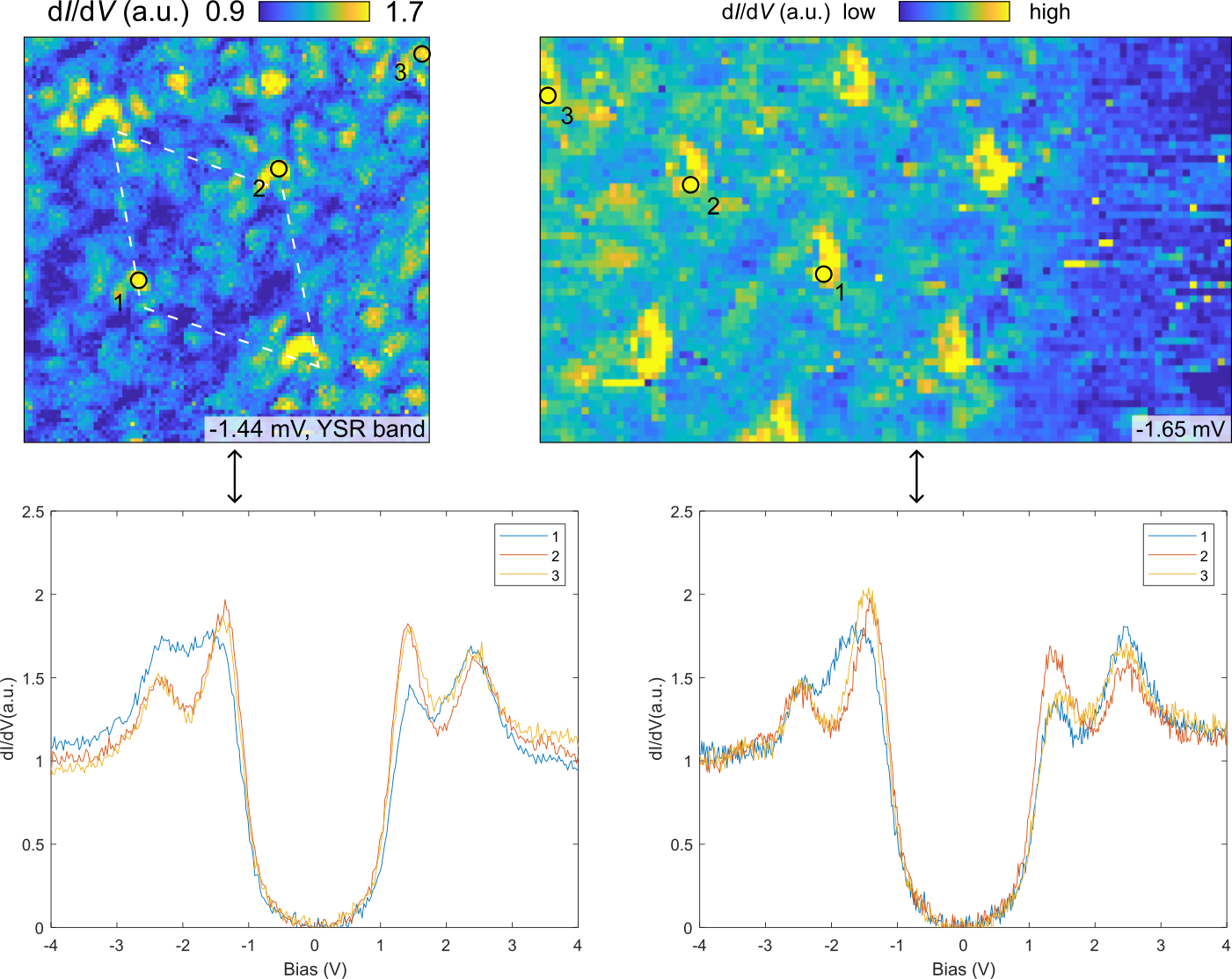}
    \caption{Real-space evolution of d$I$/d$V$ spectroscopy measured at the Ni sites of neighboring unit cells in the middle (left) and edge (right) of a SMC island.}
    \label{fig:3x3real}
\end{figure*}

In Fig.~\ref{fig:3x3real}, we show that the the spectroscopy on Ni atoms does not respect the periodicity of the 1 $\times$ 1 unit cell, consistent with the 3 $\times$ 3 reconstruction.

\section{Comparison between the measurement with a metallic and a superconducting tip}
\begin{figure*}[!h]
    \centering
    \includegraphics[width=.85\textwidth]{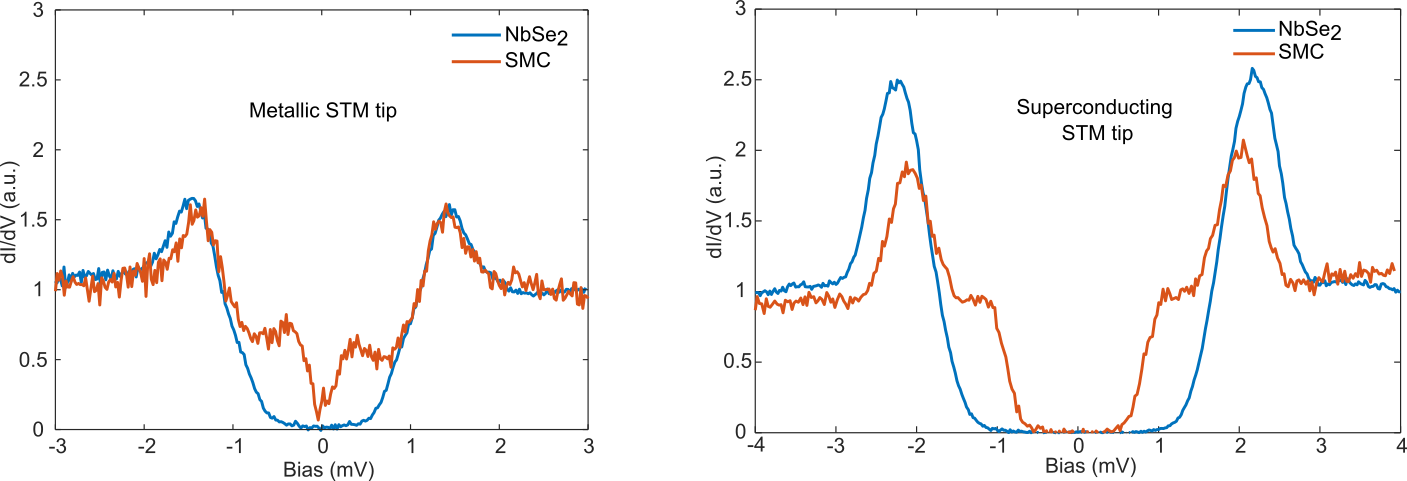}
	\caption{Measurements with a metallic (left) and superconducting (right) STM tip.}
    \label{fig:tips}
\end{figure*}

Large part of the STM experiments were done with a superconducting tip. Comparison of typical spectra acquired with metallic or superconducting tips is shown in Fig.~\ref{fig:tips}. Usually, a superconducting STM tip is used for an enhanced energy resolution in tunneling spectroscopy \cite{Pan1998,PhysRevLett.100.226801,Franke2011}. However, in our case, it was more beneficial for decreasing the interaction between the tip and the SMC due to the vdW nature of the NbSe$_2$ tip. This was realized by first having a regular metallic tip and then crashing and pulsing it on NbSe$_2$ \cite{Kezilebieke2018}. This enabled us to acquire data without destroying the measured SMC. The SMC interacts with the NbSe$_2$ substrate only with vdW forces and there are no chemical bonds within the SMC, which resulted in a metallic tip readily picking up single molecules from the SMC during a measurement.

\section{Correspondence between the STM and XAS/XMCD samples}

\begin{figure*}[!b]
    \centering
    \includegraphics[width=.9\textwidth]{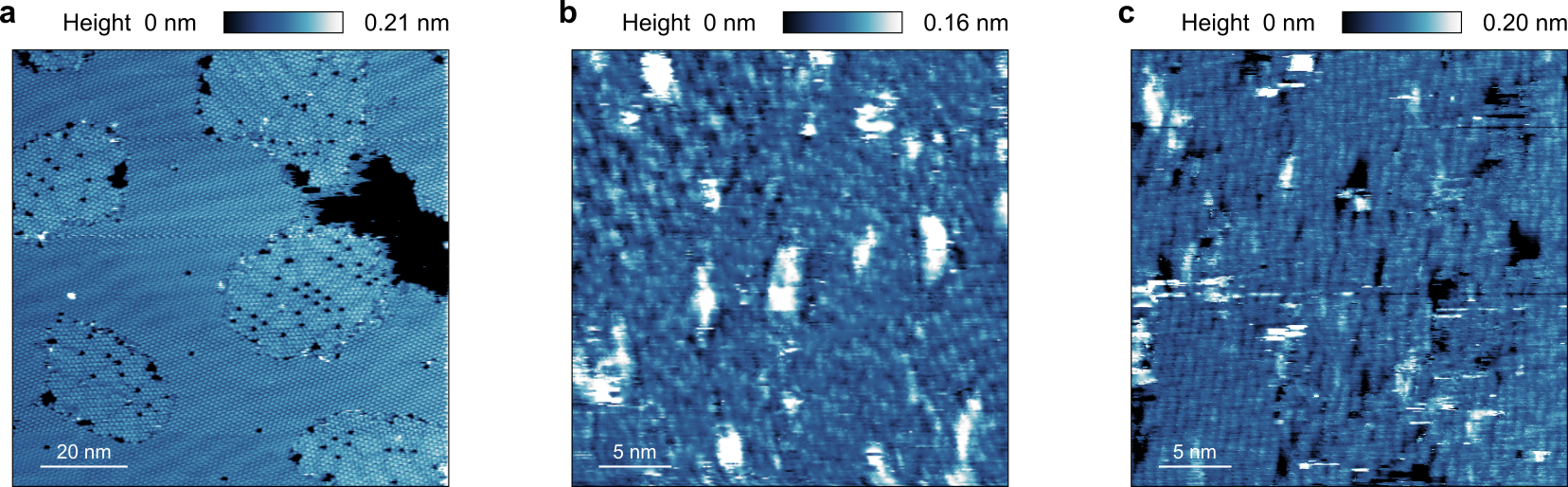}
	\caption{(a) Intermediate step of the sample growth for STM measurements. (b),(c) Growth calibration using an STM at the XMCD facility, showing the intermediate step.}
    \label{fig:samples}
\end{figure*}

\begin{figure*}[b!]
    \centering
    \includegraphics[width=.9\textwidth]{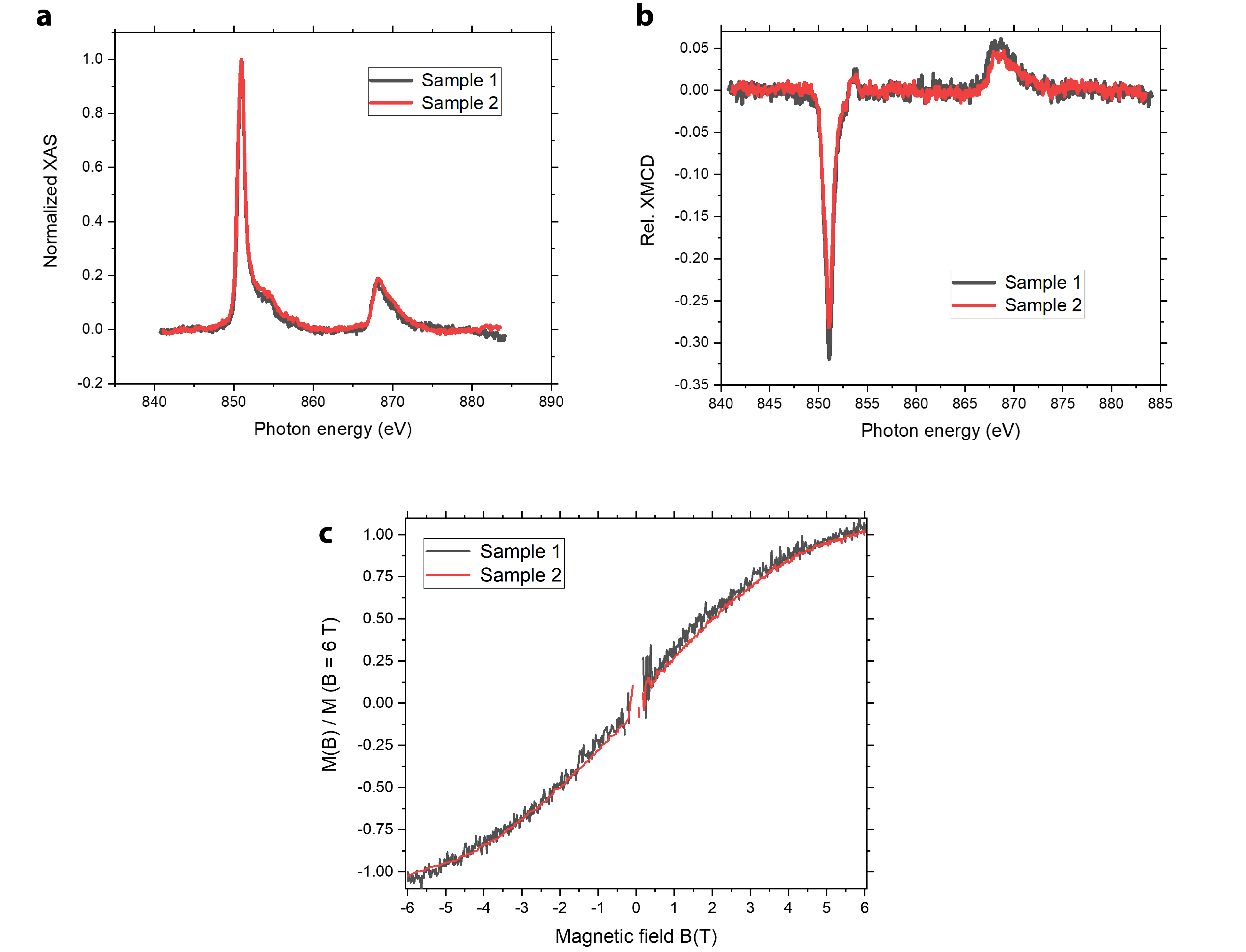}
	\caption{Comparison between NiDCA$_3$ SMC on NbSe$_2$ obtained with the same preparation method. (a) XAS, (b) XMCD, and (c) magnetization loops of NiDCA$_3$ SMC on NbSe$_2$ acquired at the nominal temperature $T_s = 2$ K and in normal incidence. Data referred to as Sample 1 are those shown in the main text, while data shown for Sample 2 are acquired on a control sample. To compare the data from the two samples, the XAS spectra have been normalized at the L$_3$ maximum. The XMCD signal are reported as the fraction of XMCD over the XAS intensity at the L$_3$. The magnetization loops are both normalized to their maximum value t $B = 6$~T.}
    \label{fig:NIDCA_comp}
\end{figure*}
The growth of the sample was carried out in three steps. The first step was a deposition of a monolayer of DCA molecules on NbSe$_2$ substrate, followed by deposition of Ni atoms and subsequent annealing of the sample that resulted in the evaporation of the excess DCA molecules. An STM image after the second step is shown in Fig.~\ref{fig:samples}a, which shows NiDCA$_3$ islands forming in a monolayer of DCA molecules. This process is very repeatable and the same growth parameters lead to the same result. Prior to the XAS and XMCD experiments, we used a molecular beam epitaxy (MBE) chamber and STM within the same UHV system to reproduce the samples. While the quality of the STM at the XMCD facility is worse (including e.g.~thermal drift) than that of the one used for STM measurements, one can use it for characterizing the sample growth. STM images of the sample after the second step are shown in Figs.~\ref{fig:samples}b,c. Fig.~\ref{fig:samples}b shows mostly a layer of DCA molecules, and a NiDCA$_3$ island in the top-left corner, while Fig.~\ref{fig:samples}c shows just a NiDCA$_3$ island. Once again, the growth was repeatable and the two scans are representative. These samples were consequently annealed at the same temperatures as the STM samples and measured.

In addition, to verify the reliability of the preparation, we also repeated the experiment on a control sample with very similar amount of Ni and with the SMC prepared in the same way as described for the sample shown in the main text. The related XAS, XMCD, and magnetization loop for this sample (sample 2) are compared to those shown in the main text (sample 1) in Fig.~\ref{fig:NIDCA_comp}. Besides a minimal difference in the XMCD amplitude, the two samples are basically identical. Notably, they both show magnetization loops that significantly deviate form the expected $S = 1$ Brillouin curve shown in the main text, further supporting the conclusion on the presence of anti-ferromagnetic interactions among the Ni centers.

\section{Angular dependent XAS measurements and details of the multiplet model}
\begin{figure*}[b!]
    \centering
    \includegraphics[width=.95\textwidth]{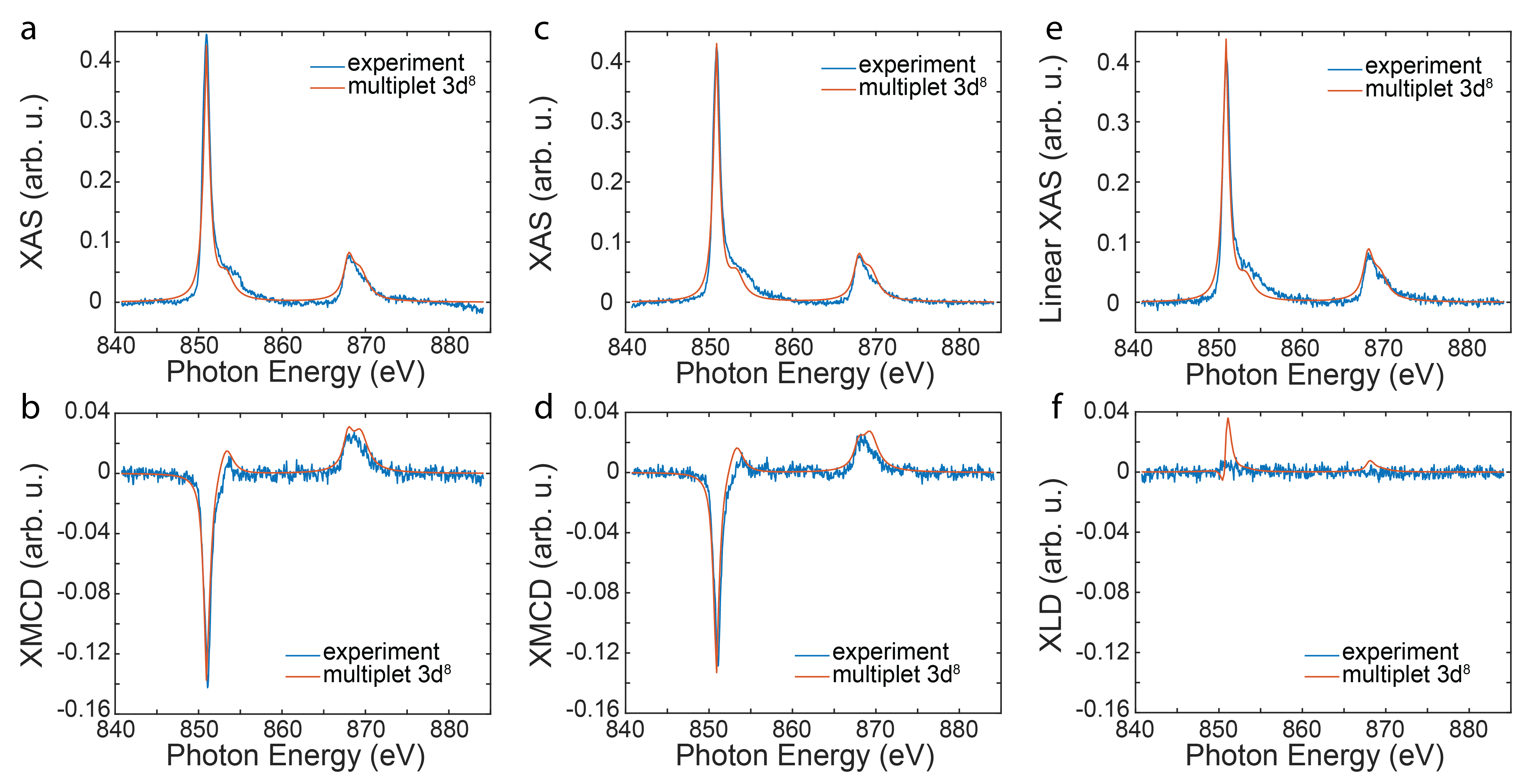}
	\caption{Angular dependent X-ray absorption measurements. (a) XAS and (b) XMCD acquired at normal incidence. (c) XAS and (d) XMCD acquired at grazing incidence. (e) linear XAS and (f) XLD acquired at grazing incidence. All measurements have been acquired at $T = 2$~K. The circularly polarized spectra have been acquired at $B = 6$~T to maximally polarize the ensemble, while the linear XAS measurements have been acquired at $B = 0.05$~T to probe the almost magnetically unperturbed electronic configuration. The best fit from the multiplet calculations are shown for comparison.}
    \label{fig:XAS}
\end{figure*}

In order to fully constrain the multiplet model and improve the accuracy of the fitting procedure, a set of circularly and linearly polarized X-ray absorption spectra were acquired at two different angles over the Ni $L_{2,3}$ edges. Fig.~\ref{fig:XAS} shows the XAS (sum of circular right (CR) and circular left (CL) polarization) and XMCD (difference between CR - CL) acquired at normal (X-ray beam and magnetic field perpendicular to the surface) and grazing incidence (beam and field at 60 degrees from the surface normal). For the latter configuration, we additionally show the summed XAS, which is the sum of the two linear horizontal (LH) and vertical (LV) polarizations defined with respect to the sample surface, together with the XLD defined as the difference between LH and LV. At high field, the system show a slightly larger XMCD along the normal incidence, indicating a small magnetic anisotropy. The very low XLD signal also indicates very little angular dependence of the absorption, pointing towards a low value of the magnetic anisotropy.

The simultaneous fit of the 6 spectra, together with the magnetization and susceptibility curves shown in Fig.~4 of the main text was performed by using 2 re-scaling parameters for the atomic Slater integrals (2p-3d and 3d-3d values), and 3 free parameters for the crystal field terms in the $C_{3v }$ symmetry (2 axial terms and 1 mixing term). In addition, we included a magnetic coupling term $J^{ex}$ through the Curie-Weiss temperature $T_{CW} = -\frac{S(S+1)}{3k_B}nJ^{ex}$ where $S$ is the spin of the Ni atom, $k_B$ is the Boltzmann constant and $n$ is the number of neighbours. The Curie-Wiess temperature is taken as a free parameter of the fit. In the regime where the measurements are performed above the ordering temperature $T_N$, the magnetic coupling affects the XMCD signal by re-normalizing the XMCD amplitude with an effective temperature $T_\mathrm{eff}=T - T_{CW}$. Based on the chosen parameters, the multiplet code generates a set of quantum states, with the lowest multiplet being a $S=1$ triplet. The separation between the singlet $m_S = 0$ and the doublet $m_S \pm 1$ defines the anisotropy parameter $D$. This parameter, together with the $T_{CW}$, defines an equivalent spin Hamiltonian over a Ni spin $i$:
\begin{equation}
H_i = D\hat{S}_{i,z}^2 + \mu_B g\hat{\mathbf{S}}_i \cdot \mathbf{B} - \frac{3k_B T_{CW}}{n S(S+1)} \hat{\mathbf{S}}_i \cdot \sum_{j\neq i}\langle \hat{\mathbf{S}}_j \rangle,
\end{equation}

where $\mu_B$ is the Bohr magneton, $g$ is the electron spin $g$-factor, and the sum is considered over the nearest neighbors $j$. The expectation value of $\langle \hat{\mathbf{S}}_j \rangle$ is obtained iteratively by calculating $\langle \hat{\mathbf{S}}_i \rangle$ and assuming $\langle \hat{\mathbf{S}}_j \rangle = \langle \hat{\mathbf{S}}_i \rangle$ at every iteration. Using this method, we calculate the expectation values of the magnetization $ M $ and magnetic susceptibility $\chi$ that we use to fit of the experiment shown in Fig.~4. The best values from the fit are shown in Table \ref{table:Parameters}.

\begin{table}[h!]
 \caption{Best parameters used for the fit of the XAS data using multiplet calculations. $D$ and $g$ are inferred from the quantum states calculated with the multiplet and are not free parameters of the fit.\label{table:Parameters}}
\begin{tabular}{||c|c|c|c|c|c|c|c||} 
 \hline\hline
 SI $2p-3d$ & SI $3d-3d$ & $E_a$ & $E_{e1}$ & $M_e$ & $T_{CW}$ & $D$ & $g$\\ 
 \hline
 0.99 & 0.90 & -7.54 eV & 4.75 eV & 15.4 eV & -2.48 K & 0.10 meV & 2.01 \\ 
 \hline\hline
\end{tabular}

\end{table}

\section{Possible origin of the 3 $\times$ 3 supercell peaks at the YSR energy}

\begin{figure*}[t!]
    \centering
    \includegraphics[width=.75\textwidth]{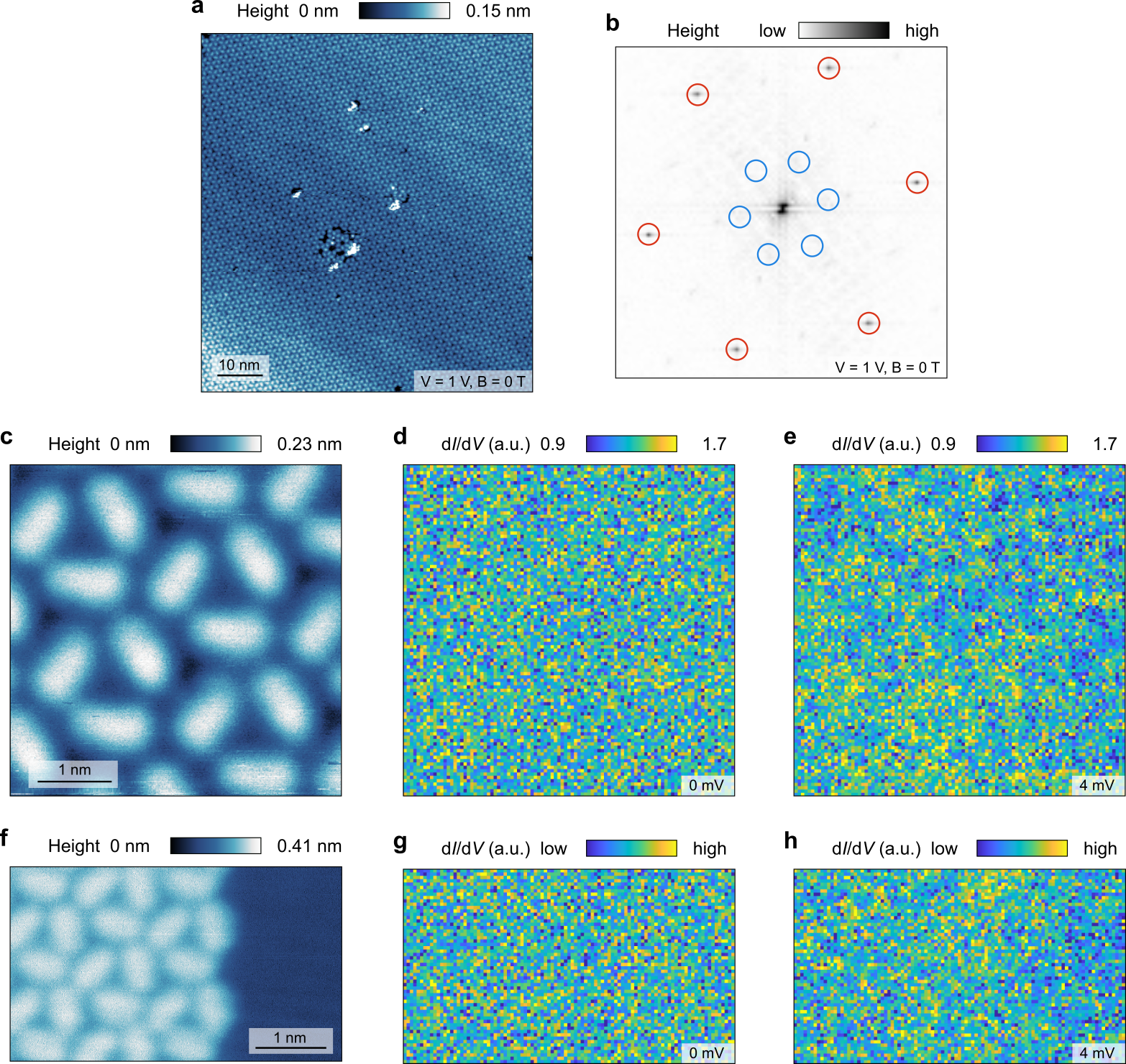}
	\caption{Absence of 3 $\times$ 3 reconstruction at energies other than the YSR band energies.}
    \label{fig:absence}
\end{figure*}

\begin{figure*}[h!]
    \centering
    \includegraphics[width=.64\textwidth]{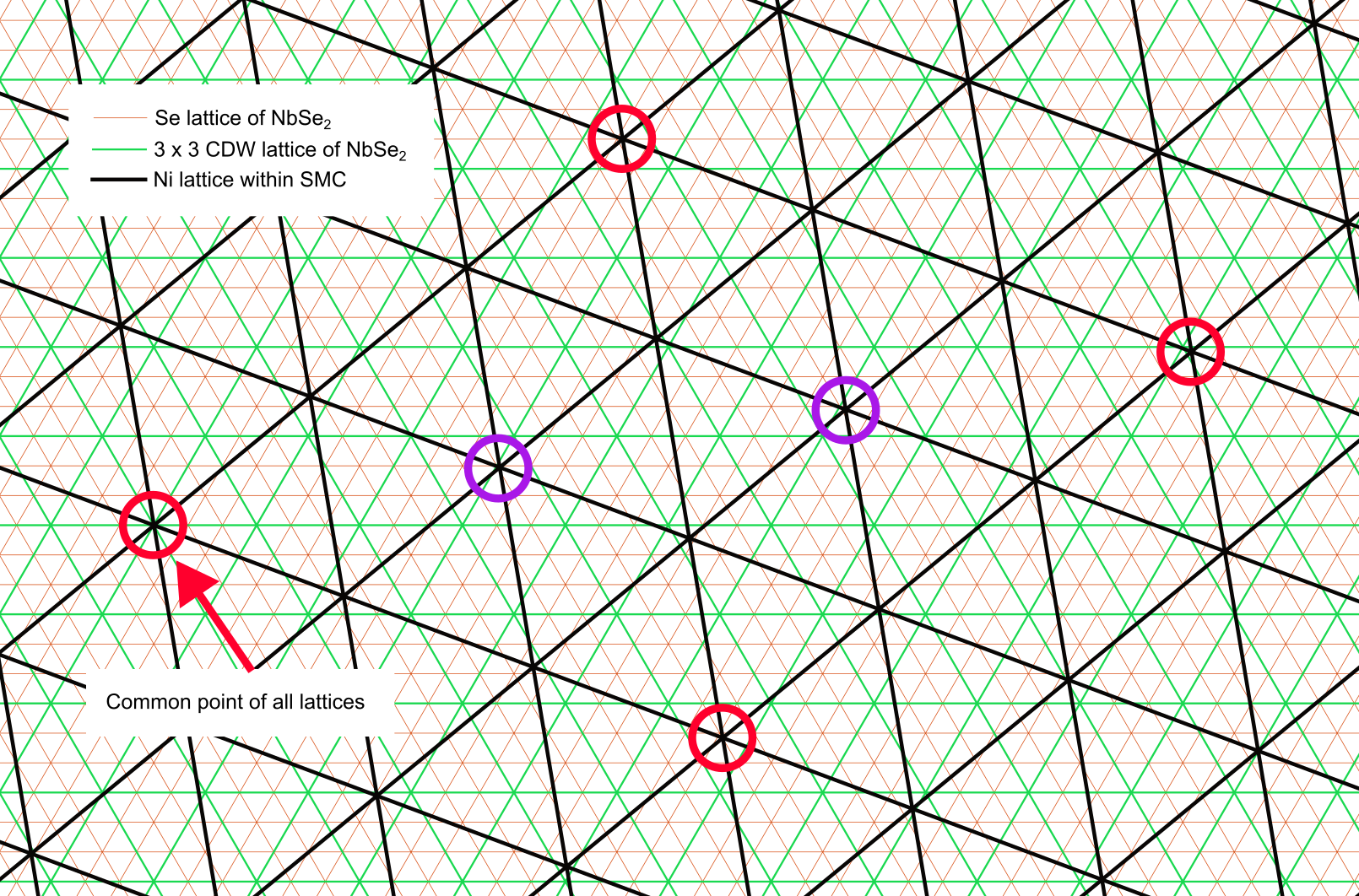}
	\caption{Overlay of three lattices: Se lattice of NbSe$_2$ (orange), 3 $\times$ 3 CDW lattice of NbSe$_2$ (green) and a Ni lattice of the SMC (black). The Ni lattice is rotated by 20.5$^\circ$ with respect to the other lattices, and the three lattices share one common point.}
    \label{fig:lattice}
\end{figure*}

An obvious candidate for the origin of the 3 $\times$ 3 supercell peaks at the YSR energy is the moir\'e pattern between the SMC Ni lattice and either the Se lattice of the NbSe$_2$ or the $3\times3$ CDW lattice of NbSe$_2$. There are several arguments why this is not the case, first of all there is no visible moir\'e pattern in STM images of the structure and the  $3\times3$  peaks appear in d$I$/d$V$($\vec{q}$,$V$) only at the energy of the YSR bands, as we show in Fig.~\ref{fig:absence}. After suppressing the superconductivity with magnetic field, the peaks vanish. Secondly, we can overlay the lattices and look at a possible moir\'e pattern. For this, we determined from our experiment the NbSe$_2$ lattice constant (0.344 nm), the SMC lattice constant (2.03 nm) and the angle between the two lattices (20.5$^\circ$), which has been consistent for all SMC islands. From the overlay, shown in Fig.~\ref{fig:lattice}, we can see that there is no $3\times3$ moir\'e pattern between the Ni and CDW of NbSe$_2$ (see red circles). We can also see that there could possibly be a moir\'e pattern between Ni and Se (see red circles), but in that case there would also be a $\sqrt{3}\times\sqrt{3}$ supercell (see purple circles), which we do not observe in our experiment. As such, we conclude that the $3\times3$ supercell observed in our experiment does not originate from a moir\'e pattern, but most probably from a magnetic $3\times3$ order of Ni.

\section{Theoretical Model}

To capture the emergence of YSR states, and their dependence on the magnetic ordering of the molecules, we consider a model for electrons in the substrate and integrating out the orbitals in the molecule.
We model the proximity between magnetic molecules to the substrate with the following tight binding model
\begin{equation}
H = \sum_{ij,s} t_{ij} c^\dagger_{i,s} c_{j,s} + 
\sum_{\alpha \in M} J \mathbf  m_\alpha \cdot  \boldsymbol{\sigma}^{s,s'} c^\dagger_{\alpha,s} c_{\alpha,s'}
+
\sum_{\alpha \in M,s} \epsilon c^\dagger_{\alpha,s} c_{\alpha,s}
+ \sum_i
\Delta c^\dagger_{i,\uparrow} 
 c^\dagger_{i,\downarrow} + h.c.
 \label{eq:tb}
\end{equation}

\begin{figure*}[t!]
    \includegraphics[width=.95\textwidth]{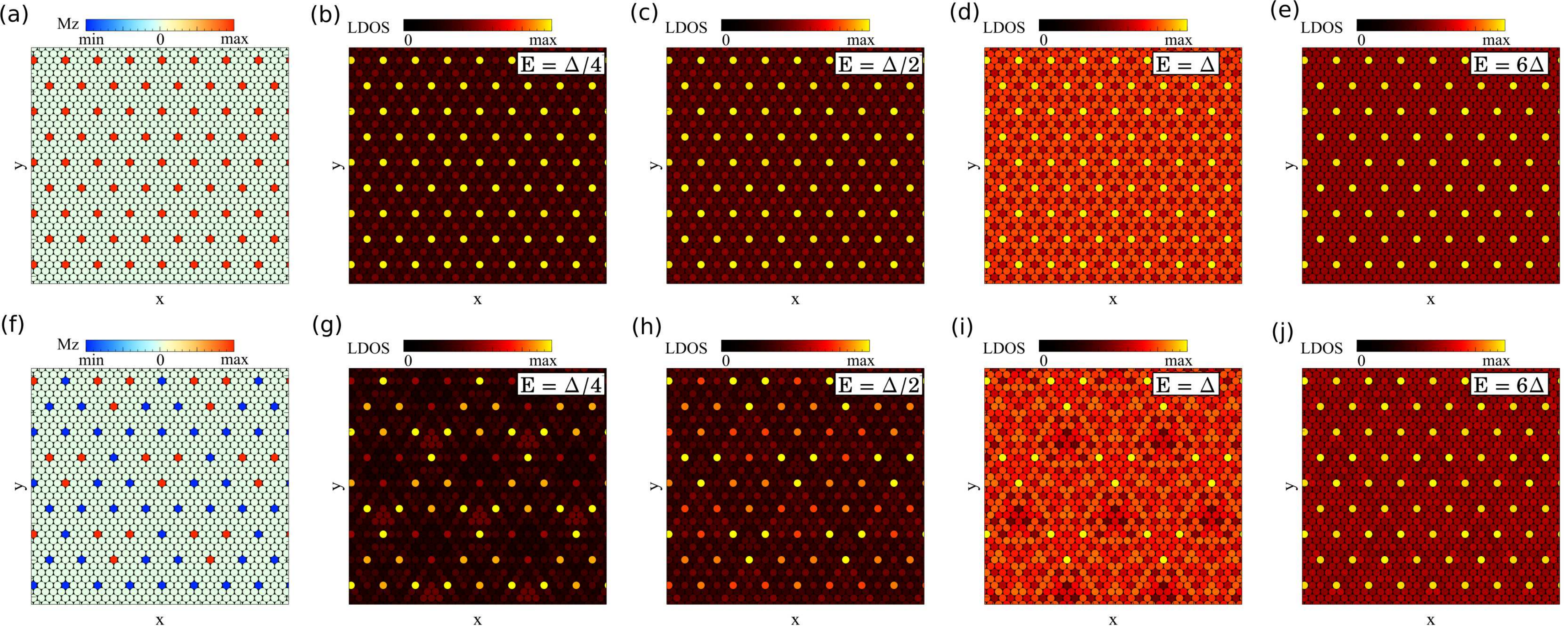}
	\caption{Superlattice with a uniform magnetic arrangement (a-e),
 and featuring the magnetic reconstruction (f-j).
 Panels (a,f) show the two magnetic arrangements,
 and panels (b-e) and (g-j) their density of states at different energies.
 In the case showing no magnetic reconstruction (a-e), the Yu-Shiba-Rusinov states
 conserve the original symmetry of the molecular lattice at all energies (b-e).
 In the presence of $3\times 3$ magnetic reconstruction, the Yu-Shiba-Rusinov
 show the magnetic reconstruction in the local density of states at energies below the gap (g-i),
 whereas recovering the full periodicity above the superconducting gap (j). We take a magnetic site every
 $4\times 4$ normal sites, $J= 7 \Delta$ and $\epsilon=0$.
    \label{fig:3x3LDOS}}
\end{figure*}
where $c^\dagger_{i,s}$ is the creation operator of electrons in the substrate,
$t_{ij}$ the electron hopping that we take as first neighbor hopping, 
$M$ the set of magnetic molecule locations and $\mathbf  m_\alpha = (0,0,\pm 1)$. The parameter $J$ captures the exchange proximity between the molecules and the electrons in the substrate, and $\epsilon$ is the local potential created by the molecules. The coupling $J$ stems from a virtual process in which an electron from the substrate tunnel to the magnetic
site, and has a similar origin as the Heisenberg coupling in conventional magnets.
We take a molecular arrangement forming a superlattice with respect to the substrate sites, and we take a fixed Ising lattice to represent a minimal solution for the $3\times 3$ reconstruction. 
Such a model builds on top of the magnetic reconstruction
of the underlying magnetic model, but it does not require knowing the original
exchange couplings in the magnet.
The term $\Delta$ accounts for the superconducting order, which for the sake of concreteness we take to be uniform in the system. As a reference, the superconducting gap of
NbSe$_2$ is on the order of 1 meV, and typical exchange couplings between magnetic molecules and
a substrate are in the range 1-10 meV. A choice of $J$ would correspond to an exchange coupling of
4 meV, consistent with the typical values expected in these systems.

With the previous model, we can show that the magnetic reconstruction is directly visualized in the YSR states as shown in Fig.~\ref{fig:3x3LDOS}. Specifically, in the absence of magnetic reconstruction, the LDOS shows the original symmetry of the molecular superlattice. Once the magnetic reconstruction is included, the YSR states reflect the new periodicity of the magnetic structure. As a reference, Fig.~4f,g in the main manuscript
are computed taking $J=4\Delta$,  an in-gap energy of $\omega = \Delta/2$,
and including a spatial smearing of the local density of states with a Gaussian local orbital of width
$0.6a$, where $a$ is the superconductor lattice constant. 

\begin{figure*}[t!]
    \includegraphics[width=0.95\textwidth]{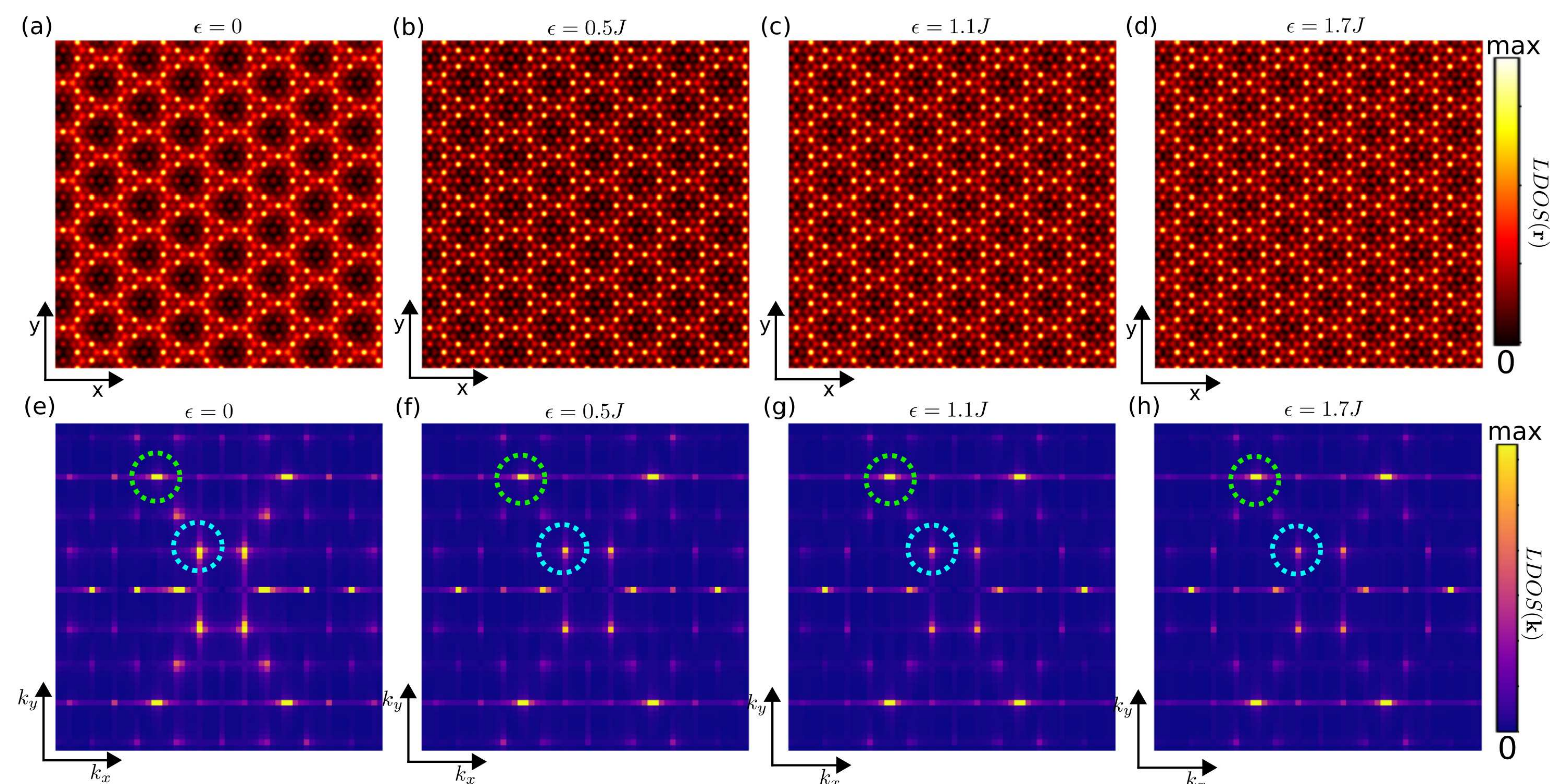}
	\caption{Real-space spectroscopy (a,b,c,d) for different values of the scalar scattering potential $\epsilon$ of the molecular
 lattice, keeping the local exchange $J$ constant. It is observed that the $3\times 3$ reconstruction appears regardless of
 the presence of the non-magnetic potential. Panels (e,f,g,h) show the Fourier transform of the real-space spectroscopy,
 showing that two peaks emerge in the different cases, the molecular lattice constant (green) and the $3\times 3$ reconstruction of the
 YSR states (cyan).
    \label{fig:pot}}
\end{figure*}

The potential scattering term $\epsilon$ in Eq.~\ref{eq:tb} does not qualitatively impact the profile of the magnetic reconstruction. This stems from the fact that in-gap YSR modes emerge due to the interplay between superconductivity and local exchange, whereas a spin-independent scattering does not create a strong impact. This can be clearly observed in Fig.~\ref{fig:pot}, where we show the simulated real-space spectroscopies and their Fourier transforms. It is observed that the $3 \times 3$ reconstruction of the YSR states appears both in the
presence and absence of the scalar scattering potential $\epsilon$. The $3 \times 3$ modulation can be observed in the Fourier transform of the real-space spectroscopy, showing that a peak associated with the periodicity of the reconstruction is present. 

\begin{figure*}[t!]
    \includegraphics[width=0.62\textwidth]{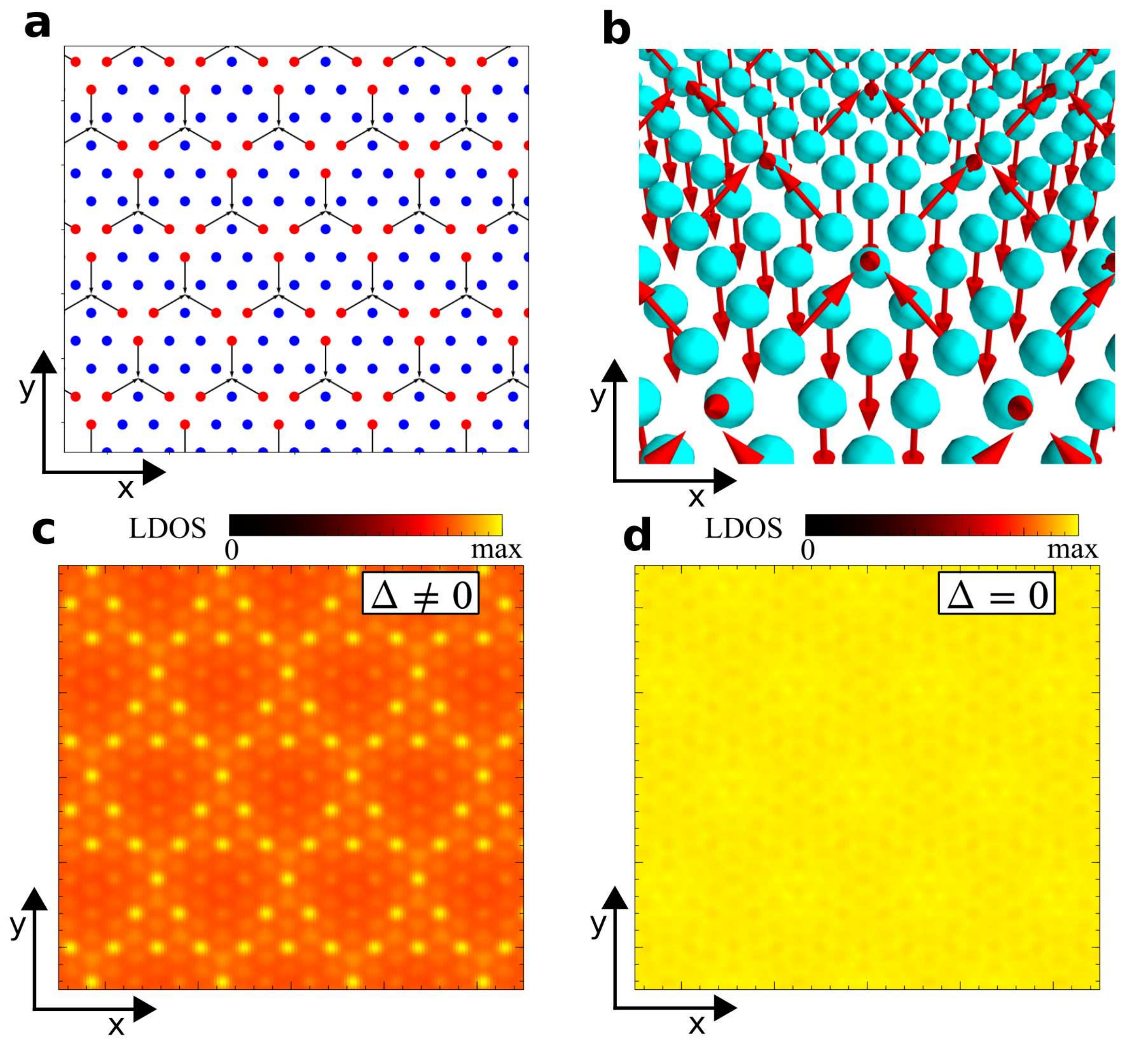}
	\caption{(a,b) Magnetic configuration of a $3\times3$
 magnetic reconstruction realizing a
 skyrmion-like arrangement. Only the sites featuring exchange
 are shown for clarity.  In the presence of $3\times 3$ magnetic reconstruction, the Yu-Shiba-Rusinov
 show the magnetic reconstruction in the local density of states in the
 presence of superconductivity (c), 
 whereas recovering the full periodicity in the absence of superconductivity (d). 
 We take a magnetic site every $4\times 4$ normal sites, and $J= 4 \Delta$.
    \label{fig:sky}}
\end{figure*}

The previous calculations consider a magnetic configuration that is collinear. In general, the ground state of the frustrated Hamiltonian is expected to show a non-collinear configuration. The sizable anisotropy observed experimentally suggests that spin-orbit coupling effects are important in the magnetic Hamiltonian. In particular, due to the mirror symmetry breaking of the surface, a Dzyaloshinskii–Moriya interaction is expected to appear. This interaction, in combination with biquadratic or ring exchange, and specially due to the geometric frustration of the lattice, can potentially lead to a non-coplanar arrangement analogous to a skyrmion (Fig.~\ref{fig:sky}a,b). This magnetic arrangement featuring a $3\times3$ periodicity of the original lattice would be directly reflected in the local density of states. Specifically, we show in Fig.~\ref{fig:sky}c,d the real-space spectroscopy in the presence and absence of superconducting order taking the $\mathbf  m_\alpha$ term in Eq.~\ref{eq:tb} 
for such a magnetic configuration. It is observed
that in the presence of superconductivity the $3\times 3$ order appears (Fig.~\ref{fig:sky}c), whereas in the absence of superconductivity the reconstruction is not observed (Fig.~\ref{fig:sky}d). The Ising magnetic configuration of Fig.~\ref{fig:3x3LDOS} can be understood as the limit in which the $m_x$ and $m_y$ components of the magnetization would be completely quenched in Fig.~\ref{fig:sky}. It is also worth emphasizing that other non-coplanar magnetic configurations featuring a $3 \times 3$ reconstruction would be consistent with the experimental data, and specifically skyrmionic configurations with an opposite winding number.
The magnetic configuration considered in our manuscript can be understood
as a minimal representation of a magnetic skyrmion,
a magnetic order that could emerge due to the competition between exchange couplings and Dzyaloshinskii-Moriya interaction
at the surface. 
These types of magnetic reconstructions have been observed experimentally in other compounds \cite{Gao2020}.
While our experiments do not allow distinguishing between collinear and non-collinear magnetic ordering,
the reconstruction observed in the YSR states shows that magnetic superstructure is appearing in the molecular complex.

The $3 \times 3$ spin reconstruction arises naturally in the presence of competing magnetic interactions.
In particular, this is a feature that has been studied extensively in skyrmions systems, where the
combination of competing interactions including 
antisymmetric exchange \cite{Dzyaloshinsky1958,PhysRev.120.91,Crpieux1998,Fert2013}, biquadratic
exchange \cite{PhysRevB.103.024439,Paul2020,PhysRevB.95.224424,PhysRevB.101.144416}, geometric frustration \cite{PhysRevLett.118.147205,PhysRevLett.108.017206,Kurumaji2019}.
For the sake of concreteness, 
we now elaborate on the mechanism that we believe is
the most relevant for our system and would give rise to such reconstructed magnetic state. Exchange interaction between molecules
is mediated by the NbSe$_2$ substrate due to the Kondo coupling between molecule and
substrate
\begin{equation}
H = \sum_{ij,s} t_{ij} c^\dagger_{i,s} c_{j,s} + 
\sum_{\alpha \in M} J \mathbf  m_\alpha \cdot  \boldsymbol{\sigma}^{s,s'} c^\dagger_{\alpha,s} c_{\alpha,s'}
+ B_z \sum_\alpha m^z_\alpha
+ ...
\label{eq:refsky}
\end{equation}
where $...$ denotes other terms in our Hamiltonian that are not required in the following discussion, and $B_z$ is the external out of plane field.
The previous Hamiltonian in the triangular lattice is known to give rise
to a variety of non-collinear magnetic phases and skyrmion states \cite{PhysRevLett.118.147205}. 
In particular, at zero external magnetic field, a
skyrmion configuration leading to a reconstruction 
was demonstrated \cite{PhysRevLett.118.147205}.
The magnetic configuration depends on the underlying dispersion
of the electron gas \cite{PhysRevLett.118.147205}, and thus a reliable
prediction of the magnetic configuration requires a full microscopic
treatment of the specific material.
Predicting thus the exact periodicity of the magnetic reconstruction
requires knowledge of the surface Fermi surface of the material, including
its possible reconstructions due to the electrostatics of the molecular lattice.
We finally note that our experiments do not allow to uniquely determine the Hamiltonian of the system,
nor infer the exact local magnetization associated to the $3 \times 3$ reconstruction. Such determination would
require spin polarized scanning tunnel microscopy measurements.

\section{DFT calculations}

\begin{figure*}[t!]
    \includegraphics[width=0.85\textwidth]{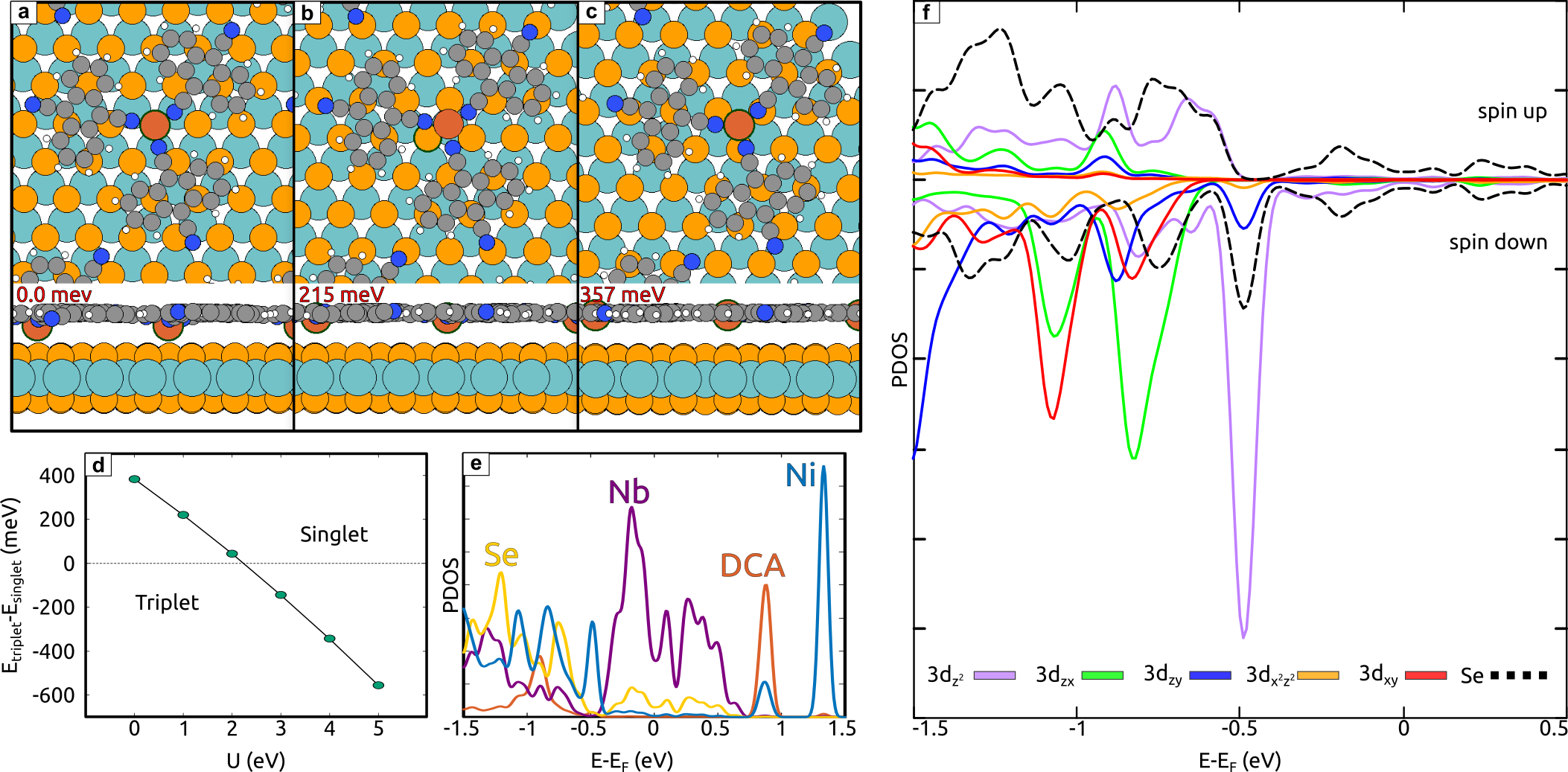}
	\caption{The structures show the geometries of the NiDCA$_3$ molecule adsorbed in three different sites of the NbSe$_2$ substrate: (a) Se site, (b) Nb site and (c) hollow site, where the energies indicated in the geometries are the energies relative to the most stable configuration (a). (d) Evolution of the spin multiplicity of the NiDCA$_3$ molecule when varying the $U$ value of the Ni atom. (e) Overall density of states projected of structure (a) on the DCA moieties and the Ni atom compared to the densities of state projected on all Nb and Se atoms that compose the substrate. (f) Projected density of states (PDOS) of the Ni 3d orbitals and the 4p orbitals of the Se atom right beneath Ni in the geometry shown in (a). The 4s orbitals of Ni and Se were omitted since they were irrelevant in the energy range where the hybridization occurs.
    \label{si_dft}}
\end{figure*}

Fig.~\ref{si_dft}a-c shows the geometry of the NiDCA$_3$ molecule in three main adsorption sites of the NbSe$_2$ substrate. The adsorption sites are defined by the position of the center Ni atom of the NiDCA$_3$, which can be sitting on top of the Se or Nb atom that composes the NbSe$_2$, as well as a hollow site meaning that no atom is located directly beneath Ni. The cell used in the calculation is equivalent to a $6\times6$ NbSe$_2$ supercell, which has a lattice parameter of 2.07 nm and can accommodate a single NiDCA$_3$ molecule with a similar orientation observed in the experiment. As shown in Fig.~\ref{si_dft}d, DFT+U with $U$ higher than 2 eV is necessary to display the triplet behavior of the Ni atom in the NiDCA$_3$ molecule observed in the experiments. Not taking $U$ into consideration drives the system to a closed shell singlet state. The density of states calculated using $U = 3$ eV (same value used throughout this work) in Fig.~\ref{si_dft}e shows that the DCA and Ni states are located well above the d-band of the NbSe$_2$ substrate. This behavior is observed regardless of the adsorption site considered, and the Ni atom shows significant hybridization with the Se states located below the Fermi level. In particular, Fig.~\ref{si_dft}f shows the PDOS of the Ni 3d orbitals and the 4p orbitals of the Se atom beneath the Ni, calculated considering the most stable energy configuration shown in Fig.~\ref{si_dft}a. A strong hybridization is observed between the spin down Se 4p orbital and the out-of-plane Ni $3d_{z^2}$ at around -0.5 eV, which gives a slight spin polarization to the Se atom. Below -0.5 eV, the spin up and down PDOS of the Se atom are nearly identical.

\bibliography{Refs}